\documentclass[a4paper,11pt]{article}
\pdfoutput=1
\usepackage{jheppub}
\usepackage{textcase}
\usepackage{graphicx,epsfig,psfrag,amssymb,hyperref}
\usepackage{multirow}
\usepackage{color,graphicx,epsfig,psfrag,amsmath,empheq}
\usepackage[dvipsnames]{xcolor}
\usepackage{bm}
\usepackage{mathrsfs,amsfonts,soul,color}
\usepackage[caption=false]{subfig}
\usepackage{hepunits}
\usepackage{enumitem}
\usepackage{placeins}
\usepackage{textcomp,upgreek}
\usepackage{tcolorbox}
\usepackage{float}
\usepackage[toc,page]{appendix}
\usepackage[ruled,vlined]{algorithm2e}
\usepackage{wrapfig}
\usepackage{hyperref}

\newcommand\snowmass{\begin{center}\rule[-0.2in]{\hsize}{0.01in}\\\rule{\hsize}{0.01in}\\
\vskip 0.1in Submitted to the  Proceedings of the US Community Study\\ 
on the Future of Particle Physics (Snowmass 2021)\\ 
\rule{\hsize}{0.01in}\\\rule[+0.2in]{\hsize}{0.01in} \end{center}}

\begin{document}

\title{Simulations of Silicon Radiation Detectors for \\ High Energy Physics Experiments}

\author[1]{B. Nachman (ed.),}
\author[2]{T. Peltola (ed.),}
\author[6, 7]{P. Asenov,}
\author[8]{M. Bomben,}
\author[5]{R. Lipton,}
\author[6, 7]{F. Moscatelli,}
\author[9]{E. A. Narayanan,}
\author[3]{F. R. Palomo,}
\author[6, 11]{D. Passeri,}
\author[9]{S. Seidel,}
\author[12]{X. Shi,}
\author[4]{J. Sonneveld}

\affiliation[1]{Physics Division, Lawrence Berkeley National Laboratory, Berkeley, CA 94720, USA}
\affiliation[2]{Department of Physics and Astronomy, Texas Tech University, Lubbock, TX 79409, USA}
\affiliation[3]{Electronic Engineering Dept., School of Engineering, University of Sevilla, 41092 Spain}
\affiliation[4]{Nikhef National Institute for Subatomic Physics, Science Park 105, 1098 XG Amsterdam, Netherlands}
\affiliation[5]{Fermilab, P.O. Box 500, Batavia Il USA}
\affiliation[6]{INFN Sezione di Perugia, Perugia, Italy}
\affiliation[7]{Consiglio Nazionale delle Ricerche - Istituto Officina dei Materiali, Perugia, Italy}
\affiliation[8]{Universit\'e de Paris \& Laboratoire Astroparticule et Cosmologie, Paris, France}
\affiliation[9]{Department of Physics and Astronomy, University of New Mexico, Albuquerque, NM 87131, USA}

\affiliation[11]{Engineering Department, University of Perugia, Perugia, Italy}
\affiliation[12]{Institute of High Energy Physics, Chinese Academy of Sciences, Beijing 100049, China}
\emailAdd{bpnachman@lbl.gov, timo.peltola@ttu.edu}

\abstract{
Silicon radiation detectors are an integral component of current and planned collider experiments in high energy physics.  Simulations of these detectors are essential for deciding operational configurations, for performing precise data analysis, and for developing future detectors.  In this white paper, we briefly review the existing tools and discuss challenges for the future that will require research and development to be able to cope with the foreseen extreme radiation environments of the High Luminosity runs of the Large Hadron Collider and future hadron colliders like  FCC-hh and SPPC.


\snowmass
}

\maketitle

\clearpage
\begin{center}
\section*{Executive Summary}
\end{center}

There are currently a variety of  tools available for simulating the properties of silicon sensors before and after irradiation.  These tools include finite element methods for device properties, dedicated annealing models, and testbeam/full detector system models.  There is no single model that can describe all of the necessary physics.  Most of these models are either fully or partially developed by HEP scientists and while there are many open-source tools, the most precise device property simulations rely on expensive, proprietary software. 

The development of these simulations happens inside existing experimental collaborations and within the RD50 Collaboration at CERN.  RD50 is essential for model research and development and provides an important forum for inter-collaboration exchange.

While existing approaches are able to describe many aspects of signal formation in silicon devices, even after irradiation and annealing, there is significant research and development (R\&D) required to improve the accuracy and precision of these models and to be able to handle new devices (e.g. for timing) and the extreme fluences of future colliders.  The US particle physics community can play a key role in this R\&D program, but it will require resources for training, software, testbeam, and personnel. 

For example, there is a great need for (1) a unified microscopic model of sensor charge collection, radiation damage, and annealing (no model can currently do all three), (2) radiation damage models (for leakage current, depletion voltage, charge collection) with uncertainties (and a database of such models), and (3) a measurement program to determine damage factors and uncertainties for particle types and energies relevant for current and future colliders.

\clearpage

\section{Introduction}

Silicon (Si) devices are now an integral component of collider-based high energy physics (HEP) experiments.  These detectors are being used as tracking detectors with precise space (and now time) measurements, for planned calorimeter detectors, and for beam monitoring.  Detailed Si simulations are essential tools for running experiments and for developing experiments of the future.  Simulations are used to model existing detectors in order to determine operational parameters like temperature and high voltage.  They are also needed to accurately describe the experiment for data analysis.  Simulations additionally are also used for device structure optimization and other detector design tasks.  Silicon simulation at the Large Hadron Collider (LHC) and future colliders requires dedicated efforts because of the unique devices needed by our community and because of the extreme radiation environments to which these devices are exposed.  A detailed review of radiation effects in the context of LHC experiments can be found at Ref.~\cite{Dawson:2764325}.

Multiple types of Si simulations are required to serve all of the needs of detector operation, analysis, and design.  At the lowest level, device simulations are used to predict the local electric field and signal formation.  Often these tools do not model the time dependence of the detector response, which is influenced by thermal annealing.  Testbeam and full detector systems also must be modeled in order to compare with data from colliders. 


At a microscopic level, the accumulation of radiation-induced defects progressively degrades various operational properties of a silicon sensor both inside the Si-bulk and at the surface on the Si/SiO$_2$-interface.  These properties are presently addressed by several defect models that apply 2--5 acceptor and donor type trap levels in the Si-substrate, and fixed oxide charge density ($N_\textrm{f}$) with/without 1--2 acceptor and donor type interface-trap ($N_\textrm{it}$) levels at the surface.  At a macroscopic level, silicon devices are characterized by their signal formation, leakage current, depletion voltage, etc.  The signal formation is then responsible for higher-level quantities like track reconstruction and physics analysis with charged-particle tracks.


The goal of this report is to consider Si simulations in the context of the US Community Study on the Future of Particle Physics (Snowmass 2021).  The following discussion is divided into two parts, introduction of existing simulation tools and radiation defect models and then the challenges faced in their further development for 
future requirements. 
In the first part, after describing the mathematical models applied in the numerical Technology Computer-Aided Design (TCAD)-simulations in Section~\ref{TCADmathModels}, 
the currently utilized tools for the per-sensor simulations, as well as for the testbeam-telescope-setup and full-detector-system simulations are introduced in Section~\ref{Tools}. 
Also presented in Section~\ref{Tools}, and based largely on the results summarized in the Chapter 7 of the recent CERN Yellow Report~\cite{Dawson:2764325}, are the various radiation defect models for bulk 
and surface damage or their combination, that have been developed in the experiments of the LHC. Additionally, speedup investigations of large-mesh-size ($\sim10^5-10^6$ 
mesh-points) TCAD devices and TCAD's process simulation option are discussed in Sections~\ref{LargeMesh} and~\ref{Process_TCAD}, respectively. The second part of the paper in 
Section~\ref{CandN} addresses both the observed and anticipated requirements and challenges on the simulation frameworks and the further development of radiation defect models toward 
more unified approaches that are able to account for the extreme radiation environments in the future colliders. Potential issues with proprietary software and computational demands 
of large mesh-size devices are discussed in Sections~\ref{PropSoftware} and~\ref{ComputTCAD}, respectively, while the development needs of simulation chains between packages and the 
sources of uncertainties to be taken into account in the TCAD modeling are addressed in Sections~\ref{TCAD_TCT} and~\ref{uncertTCAD}, respectively. Finally, the approaches to upgrading 
and unifying existing defect models in the radiation fluence region beyond $1\times10^{16}$ n$_{\textrm{\small eq}}/$cm$^{2}$, expected for future colliders, are considered in 
Sections~\ref{unifiedTCAD},~\ref{Improve_HM} and~\ref{TCADfuture}.

\section{Existing Tools}\label{Tools}



\subsection{TCAD Simulations for Detector Properties}\label{TCAD}

Radiation levels above $\sim10^{13}~\textrm{n}_\textrm{eq}/\textrm{cm}^{2}$ introduce observable damage to the crystal structure of a silicon sensor. Fluences
beyond $1\times10^{14}~\textrm{n}_\textrm{eq}/\textrm{cm}^{2}$ lead eventually to a significant degradation of the detector performance. 
In the immense radiation environment of the HL-LHC, defects are introduced both in the silicon substrate (bulk or displacement damage) and in
the SiO$_2$ passivation layer, that affect the sensor performance  
via the interface with the silicon (surface damage). The multitude of observed defect levels  
after irradiations with hadrons or higher energy leptons \cite{Radu2015}, set up a broad parameter space that is not  
practical to model and tune. Thus, minimized sets of defects constituting various effective defect models have been applied as the approach for the simulations of irradiated silicon detectors. Most of these defect models have been developed for the commercial finite-element
TCAD software\footnote{See Appendix~\ref{TCADmathModels} for the governing equations.} frameworks Synopsys Sentaurus\texttrademark\footnote{http://www.synopsys.com}
and Silvaco Atlas\texttrademark\footnote{http://www.silvaco.com}.

\begin{figure*}[htb]
     \centering
     \subfloat[]{\includegraphics[width=.48\textwidth]{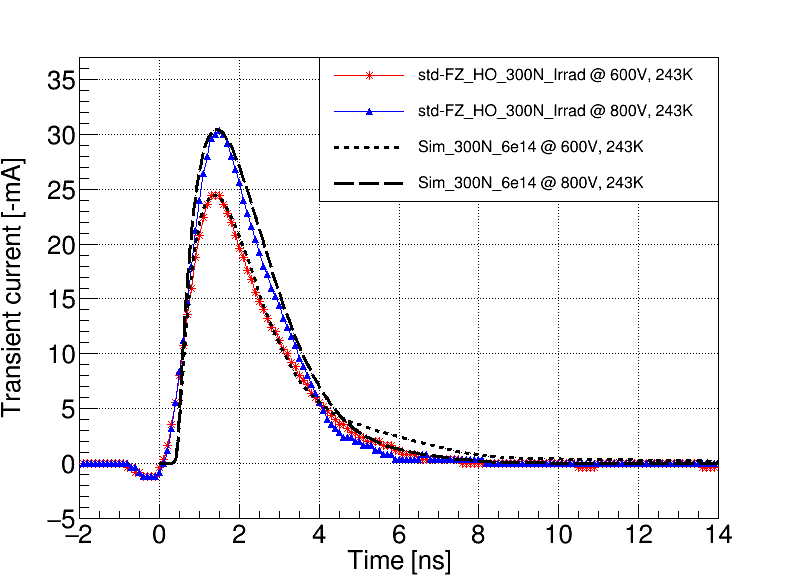}\label{TCT_Irrad}}
     \subfloat[]{\includegraphics[width=.53\textwidth]{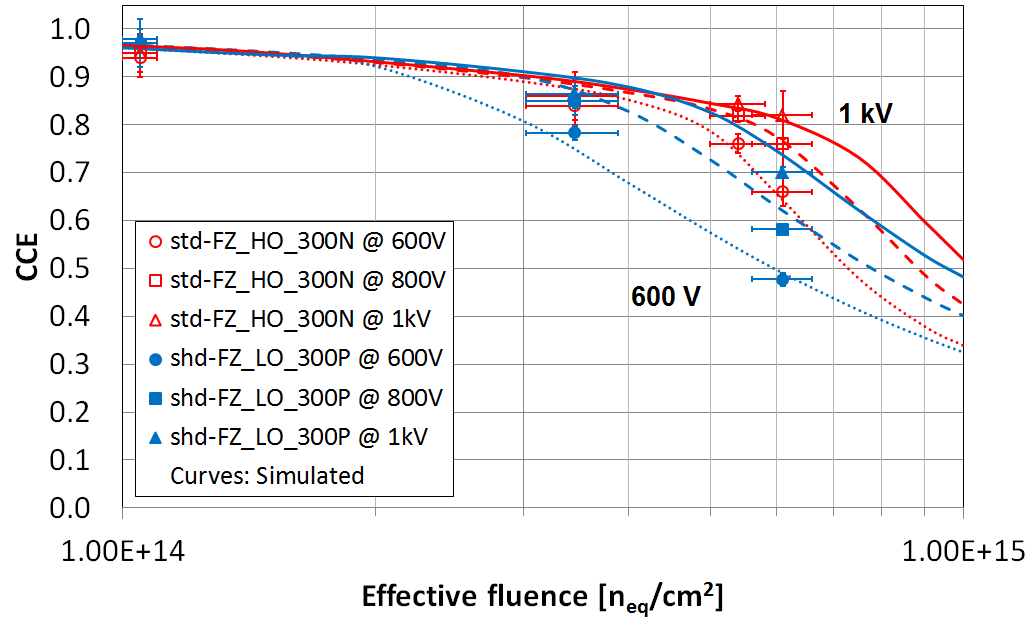}\label{CCE_F300}}
    \caption{Examples of bulk-damage modeling \cite{Peltola2020}. (a) Measured and simulated IR-laser induced transient signals for 
    300-$\upmu$m-thick $p$-on-$n$ pad-sensor (300N) after neutron irradiation to the fluence of $(6.1\pm0.5)\times10^{14}~\textrm{n}_\textrm{eq}/\textrm{cm}^{2}$. The simulation applied neutron defect model 
    \cite{Eber2013} with a fluence of $6.0\times10^{14}~\textrm{n}_\textrm{eq}/\textrm{cm}^{2}$. 
    The sensor parameters used in the simulation were extracted from $CV/IV$-measurements. (b) Measured and simulated evolution of charge collection efficiency (CCE) with voltage and 1-MeV equivalent neutron fluence for 300-$\upmu\textrm{m}$ thick sensors at -30 $^{\circ}$C.   
Simulated results use dotted, dashed and solid curves for 600 V, 800 V and 1 kV, respectively. 
} 
\label{fig:fig13}
\end{figure*}

The modeling of the main macroscopic effects of the bulk damage on high-resistivity Si-sensors irradiated by hadrons--the change of the effective doping concentration $N_\textrm{eff}$ (resulting in modified electric field distribution and change of operational voltage), the increase in the leakage current proportional to the fluence and the degradation of charge collection efficiency (CCE)--includes two-level \cite{Eremin2002,Chiochia2005,Eremin2011,Verbit2012,Eber2013}, three-level \cite{Petasecca2006,Pennicard2008,Passeri2016,FOLKESTAD201794}, 
as well as a five-level models~\cite{schwandt2019new}. Figure~\ref{fig:fig13} demonstrates a comparison between measured transient signals (Figure~\ref{TCT_Irrad}) and CCE (Figure~\ref{CCE_F300}) of neutron irradiated sensors and simulated results produced by a two-level defect model for neutrons. 

\begin{figure*}
     \centering
     \subfloat[]{\includegraphics[width=.477\textwidth]{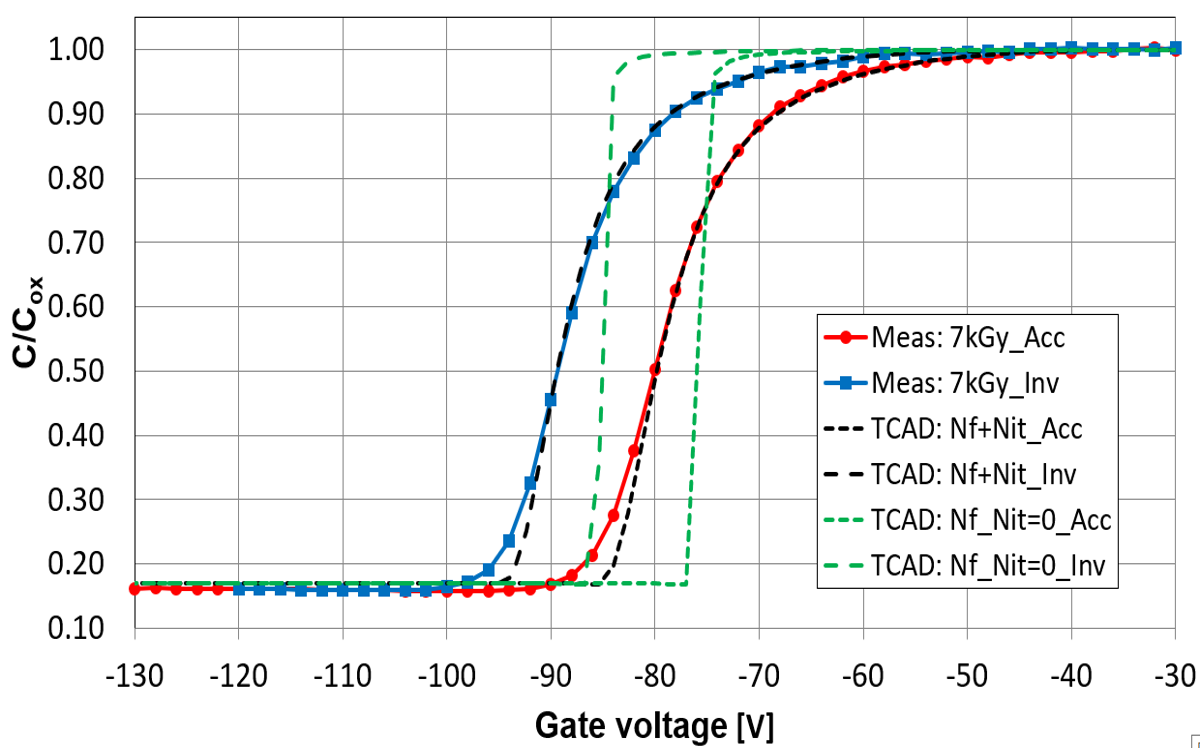}\label{nMOS_7kGy}}
     \subfloat[]{\includegraphics[width=.48\textwidth]{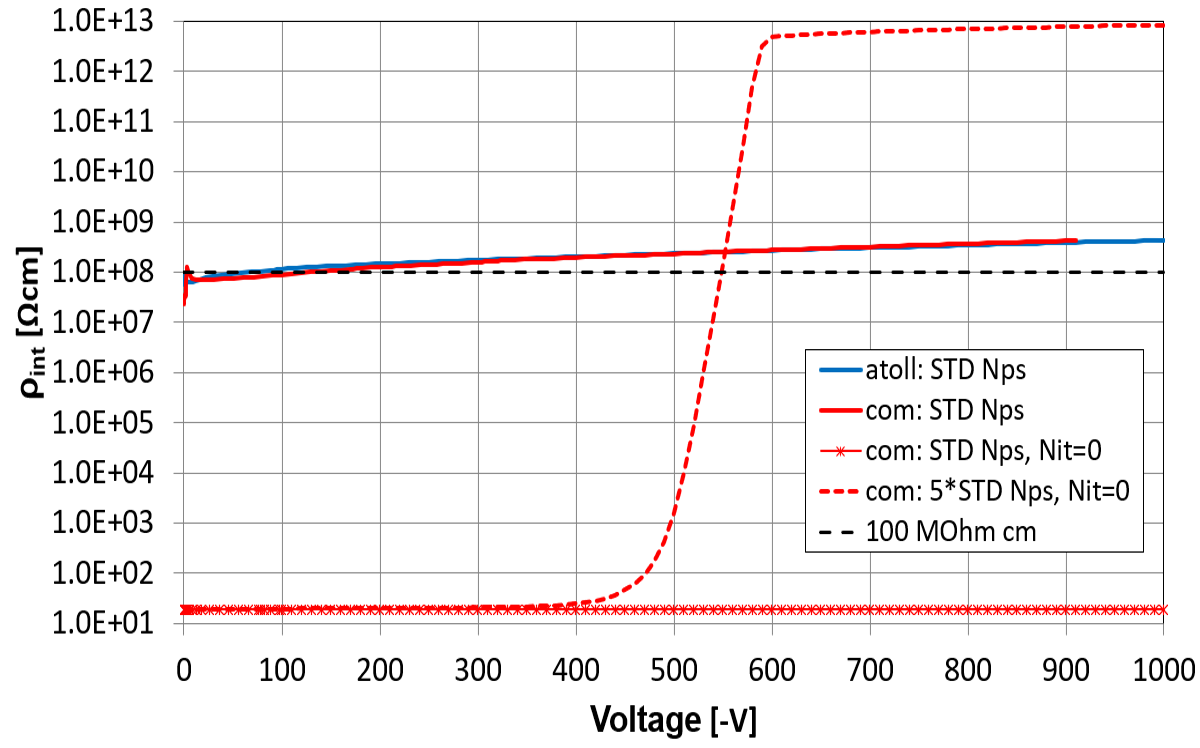}\label{Rint_Nit}}
    \caption{Examples of surface-damage modeling. (a) Measured and simulated $CV$-curves at $T=\textrm{RT}$ and $f=2$ kHz of a $\gamma$-irradiated 200-$\upmu$m-thick $n$-bulk MOS-capacitor with $t_\textrm{ox}=700$ nm for the dose of $7.0\pm0.4$ kGy. Measurements included voltages starting from both inversion (Inv.) and accumulation (Acc.) regions. (b) Inter-pad resistivity at -20$^{\circ}$ C for individual (atoll) and common p-stop isolation implants with HGCAL-parameters, simulated by applying $N_\textrm{f}$ and $N_\textrm{it}$-parameters tuned from MOS-capacitor measurements and simulations for the dose of 23.5 kGy. The 100 M$\Omega\cdot$cm (black dashed line) represents conservative estimation for high level of pad isolation. STD $N_\textrm{ps}$ = standard value of p-stop peak doping.
   } 
\label{fig_CCE300}
\end{figure*}

Simulating the effects of the surface damage on Si/SiO$_2$-interface irradiated by charged hadrons, gammas, or X-rays --like the change of inter-electrode resistance ($R_\textrm{int}$) and 
capacitance ($C_\textrm{int}$), modified electric fields at implant edges and charge-injection position dependence of CCE (CCE($x$))-- involves approaches, where surface damage is approximated 
solely by the fixed oxide charge density ($N_\textrm{f}$) \cite{Verzellesi2000,Piemonte2006,Unno2011} or by three interface traps ($N_\textrm{it}$) with parameters matching the 
measurements of X-ray irradiated MOS-capacitors \cite{Zhang2012}. Results from a further model that combines the two approaches is presented in Figure~\ref{fig_CCE300}, where the 
characteristics of the $CV$-curves of a $\gamma$-irradiated MOS-capacitor are only reproduced by simulation by including in addition to $N_\textrm{f}$ both acceptor and donor-type 
deep $N_\textrm{it}$ ($N_\textrm{it,acc,don}$) at the Si/SiO$_2$-interface (Figure~\ref{nMOS_7kGy}). By using both $N_\textrm{f}$ and $N_\textrm{it,acc,don}$ as an input for 
an $R_\textrm{int}$-simulation with typical CMS High Granularity Calorimeter (HGCAL) isolation implant parameters in Figure~\ref{Rint_Nit}, the experimentally observed 
$R_\textrm{int}$-values are quantitatively reproduced (atoll and common p-stops with STD $N_\textrm{ps}$). By excluding $N_\textrm{it,acc,don}$ from the simulation, the pads 
become either shorted (STD $N_\textrm{ps}$, $N_\textrm{it}=0$) for all voltages or reach isolation only above 500 V of reverse bias voltage for an extreme value of p-stop peak 
doping ($5\times\textrm{STD}$ $N_\textrm{ps}$, $N_\textrm{it}=0$).

First steps towards a unified bulk and surface defect model include an approach where two bulk defect levels are augmented by an acceptor-type $N_\textrm{it}$ with 2 $\upmu$m 
depth distribution from the surface (Sentaurus \cite{Peltola2014,Peltola2015}), and another approach where two bulk defect levels are complemented by two acceptor-type 
$N_\textrm{it}$ (Silvaco \cite{Dalal2014b}). More details on the development of the defect models are described in e.g., \cite{Peltola2015r}.

Another set of approaches to unify the surface and bulk defect models with common characteristics is known as the Perugia model (named after the city of the group which developed it). It takes into account the fact that in principle, the radiation damage in silicon sensors is caused by two different factors: the ionizing energy loss and the non-ionizing energy loss. Although the ionizing energy loss is important for the signal formation and is usually reversible, it can cause irreversible effects on the oxide by introducing positive oxide charge in the SiO$_2$, by increasing the number of bulk oxide traps and by increasing the number of interface traps. These effects are referred to as surface damage in the model, and they influence the operation of segmented silicon sensors with respect to the inter-electrode isolation, the breakdown voltage and the charge collection efficiency in real-life experiments. The non-ionizing energy loss is responsible for the introduction of defects into the silicon lattice through the displacement of crystal atoms, usually produced due to the impact of high-momentum particles. These impacts lead to point and cluster defect generation and hence to the introduction of deep-level trap states which act like generation-recombination centers. Non-ionizing effects are referred to as bulk damage in the Perugia model and on a macroscopic scale they are responsible for the increase of the leakage current in silicon sensors, the changes in the effective space charge concentration and the charge collection efficiency~\cite{Moscatelli:2017sps}.

\begin{figure*}
     \centering
      \includegraphics[width=.5\textwidth]{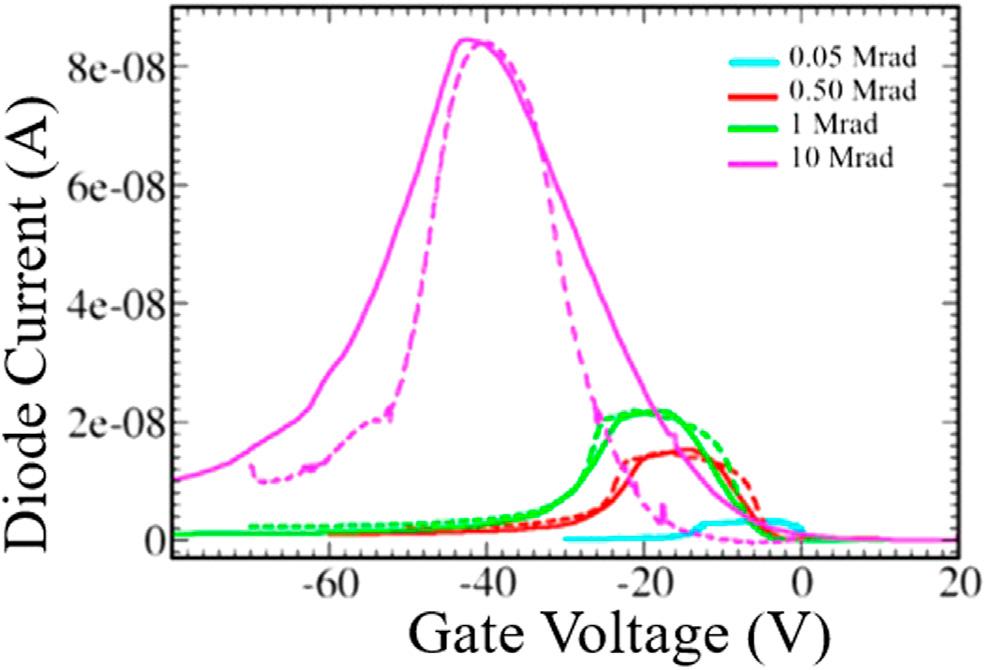}
    \caption{$I$-$V$ characteristics of $n$-on-$p$ HPK GCD: comparison between the measured and simulated diode current as a function of gate voltage (with $p$-spray isolation layer). Measurements in solid line and simulations in dashed line. The model used in the simulation is \textit{Perugia surface model 2019}. 
    } 
\label{GCD_IV_PER}
\end{figure*}

\begin{figure*}
     \centering
      \subfloat[ ]{
      \includegraphics[width=.49\textwidth]{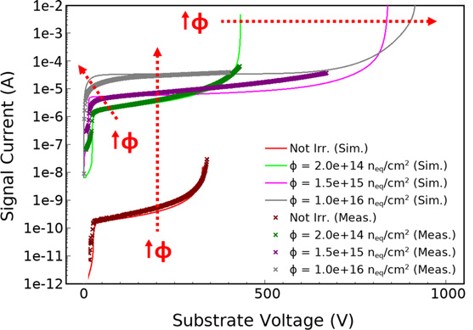}}
       \subfloat[ ]{
      \includegraphics[width=.49\textwidth]{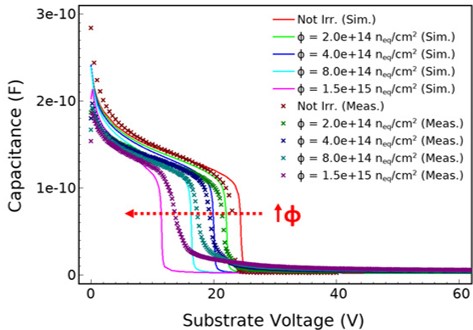}}
    \caption{(a) $I$-$V$ 
    and (b) $C$-$V$ 
    simulated curves compared with experimental data before and after irradiation for a Low-Gain Avalanche Diode (LGAD) produced by FBK. The sensor area is $1\,$mm$^2$, the thickness is 55$\,\upmu$m and the temperature is 300$\,$K. In the simulation, the \textit{New University of Perugia TCAD model} was coupled with an analytic model developed by the Torino group that describes the mechanism of acceptor removal in the multiplication layer.
    } 
\label{CV_IV_PERUGIA}
\end{figure*}

In the original Perugia model \cite{Passeri2016,Moscatelli:2017sps}, a three level model is presented for simulating the bulk damage effects for $n$-type and $p$-type substrates. In order to incorporate also the surface damage effects an extension of the model has been made referred to as the \textit{Perugia surface model 2019}~\cite{Morozzi:2020yxm,Morozzi:2021kwo}. The surface damage effects can be mainly described by three parameters: the oxide charge density ($N_\textrm{ox}$) 
and the two interface trap states ($N_\textrm{it}$) for donors and acceptors. The values of the above quantities can be extracted from high-frequency and quasi-static $CV$-measurements and then used as inputs in the simulation. Macroscopic factors, such as the capacitance or the leakage current, are calculated as outputs. This version is able to reproduce the radiation damage macroscopic effects up to the order of $10^{15}\,$n$_\textrm{eq}$/cm$^2 \,$ 1-MeV neutron equivalent fluences and X-ray photon doses of up to 10$\,$Mrad. The \textit{Perugia surface model 2019} damage modeling scheme, is fully implemented within the Sentaurus TCAD environment. Its main advantages are the dependence on  a limited number of parameter relevant for physics and the versatility of the simulation approach. Up to now, it has already been compared with experimental measurements performed on Gate-Controlled Diodes (GCD) (Figure~\ref{GCD_IV_PER}) and Metal-Oxide-Semiconductor (MOS) capacitors at Fondazione Bruno Kessler (FBK) in Italy, Hamamatsu Photonics K.K. (HPK) in Japan and Infineon Technologies (IFX) in Austria on both $n$-type and $p$-type substrates. Furthermore, an updated radiation damage model (called \textit{New University of Perugia TCAD model}) has been fully implemented within the simulation environment, to have a predictive insight into the electrical behavior and the charge collection properties of Low Gain Avalanche Detectors (LGAD) detectors. By coupling the \textit{New University of Perugia TCAD model} with an analytical model that describes the mechanism of acceptor removal in the multiplication layer of the LGAD, it has been possible to reproduce experimental data with high accuracy (Figure~\ref{CV_IV_PERUGIA}) \cite{Croci:2022ygd}. Thanks to the intrinsic multiplication of the charge within the LGADs, it is possible to improve the signal-to-noise ratio (SNR), thus limiting its drastic reduction with fluence, and hence a special attention has been given to the choice of the avalanche model, which allows the simulation findings to better fit with experimental data. The Massey avalanche model was chosen as the best option since it is the only one that does not underestimate the breakdown voltage of the device and leads to successful comparison with the experimental data \cite{Massey}.

This last version of the Perugia model is able to reproduce the behavior of 3D-devices as well.

\subsubsection{TCAD Alternatives}
There are simpler software alternatives to TCAD like TRACS, \cite{tracs}, Weightfield2, \cite{weightfield} or KDetSim\footnote{http://kdetsim.org} \cite{kdetsim1}. All of them solve the Poisson equation (\ref{eq:poissoneq}) for a charge distribution $N_\textrm{eff}$. Instead of calculating $N_\textrm{eff}$ from the full set of continuity equations, carrier drift is done in a static electric field and calculated by steps and the detector signal (induced current) comes from the Shockley-Ramo theorem, \cite{shockley, ramo}:
\begin{equation}
I_i = -q \overrightarrow{v}(x)\cdot \overrightarrow{E}_w (x)\,,
\label{eq:ramo}
\end{equation}
where $x$ is the carrier position, $v(x)$ is its drift velocity that depends on the applied electric field $E(x)$ and $E_w(x)$ is the weighting field for the particular $i$-electrode. This lightweight approach does not substitute a TCAD software but it is appropriate, for example, to design a full TCAD simulation and for obtaining approximate results much faster than with any TCAD suite, which would be more precise and capable but also more demanding in computational resources. In any case, none of the lightweight alternatives are currently suited to calculate trap dynamics.
\subsubsection{Simulations of Large-Mesh Devices}\label{LargeMesh}
\begin{figure}
\centering
\subfloat[]{\includegraphics[width=.36\textwidth]{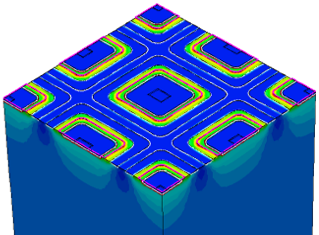}\label{fig:3Dpix}}
\subfloat[]{\includegraphics[width=.36\textwidth]{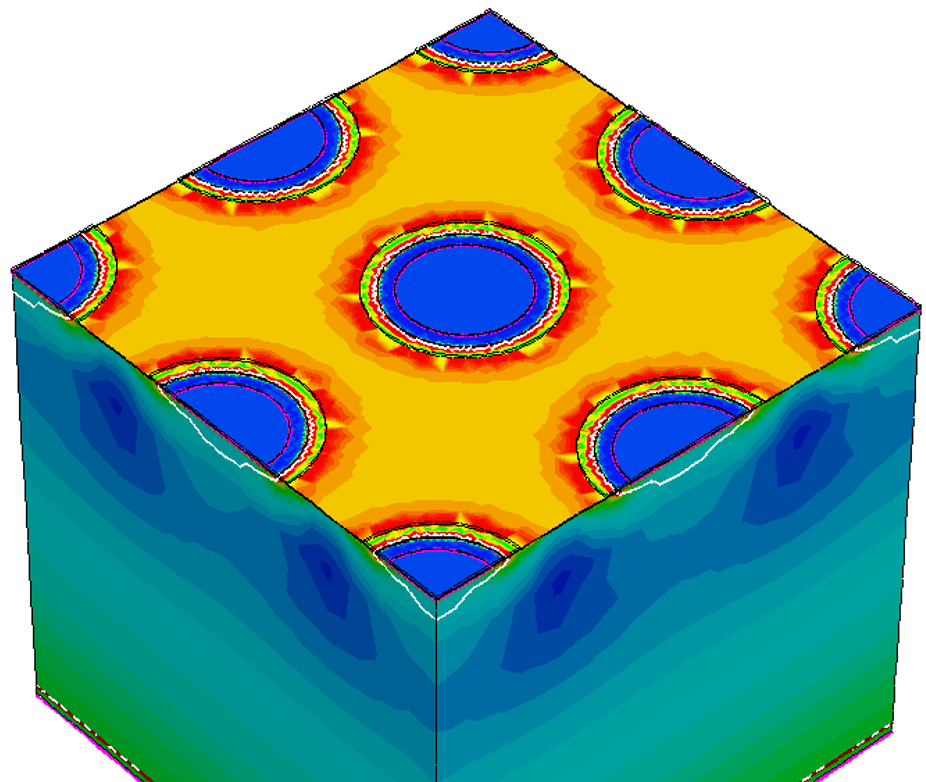}\label{fig:MediPix}}
\subfloat[]{\includegraphics[width=.3\textwidth]{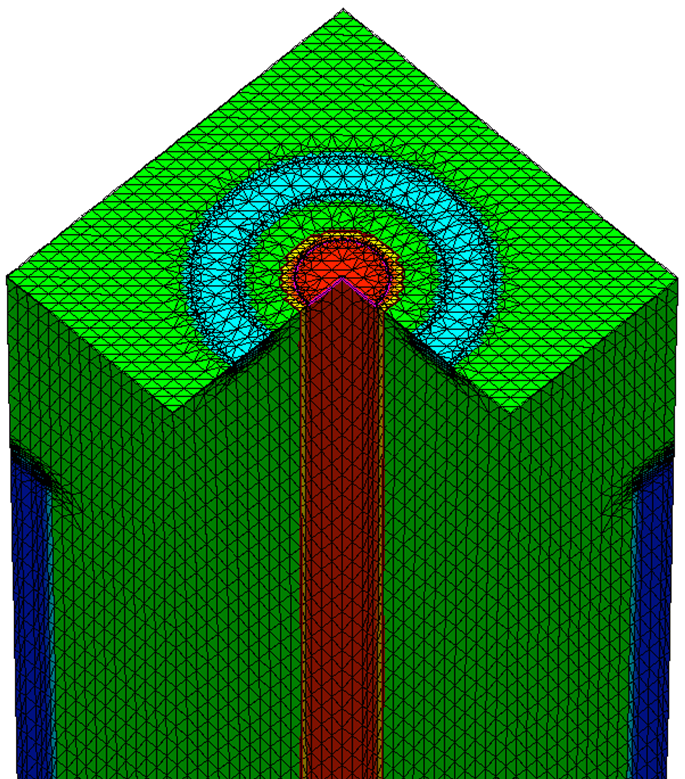}\label{fig:3Dcolumn}}
\caption{3D-sensor structures designed in Sentaurus TCAD. (a) Electric fields in a $50\times50$ $\mu\textrm{m}^2$ $n$-on-$p$ pixel sensor for CMS Tracker. 
(b) Electric fields in a $p$-on-$p$ MediPix sensor \cite{Peltola2016p} with a 55 $\mu\textrm{m}$ pitch. (c) Sliced view of a double column-double sided 3D $n$-on-$p$ sensor with a p-stop isolation. Aluminum and oxide have been stripped from the surface.} 
\label{fig:12b_3D}
\end{figure}

In the Synopsys Sentaurus TCAD software framework, the device structures can be generated in both 2D or 3D. Sensors that have negligible contribution to the weighting field from the third dimension, i.e. diodes and strip sensors, can be
accurately modeled in 2D, and extended to the dimensions of a real device by an appropriate factor. The 3D device structures, although requiring much computing time and processing capacity, are mandatory for reliable simulations of the pixel and 3D-columnar sensors, presented in Figure~\ref{fig:12b_3D}.
In a typical parameter scan the number of nodes in the simulation can
be 10--20, while the simulation of a single node for a 3D-structure with large enough mesh-size to reproduce a realistic device (e.g. smooth implant shapes at the electrodes) can be in the order of days in a standard multi-core processor. Therefore, considering different options for the optimization of the simulation times is highly motivated. 

\begin{figure*}
     \centering
      \includegraphics[width=.7\textwidth]{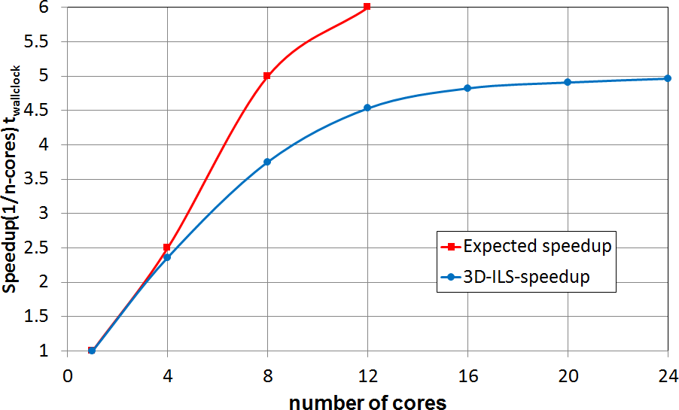}
    \caption{Speedup by shared-memory parallelization in a 3D-detector structure with about $10^6$ mesh-points, where the physics equations are solved. Theoretical speedup courtesy of Synopsys support.
    } 
\label{ILS1}
\end{figure*}

Speedup investigation of a 3D-device with about $10^6$ mesh-points in Figure~\ref{ILS1} shows, that shared-memory parallelization benefits simulation execution time by a factor of five when the number of threads is increased from one to match the number of physical cores in the processor. A further factor of about three is gained when the computation method of the physics equations is changed from a direct linear solver (optimal for 2D-simulations) to an iterative linear solver (ILS) with tuned parameters to avoid compromising the accuracy of the simulation results. Weakly supported distributed processing (cluster computing) in Synopsys TCAD is possible to be compensated in a cluster by generating copies of the simulation project that are dedicated for each node of the parameter space and running these simultaneously. 
%

\subsection{Process Simulations for Device Development}\label{Process_TCAD} 
\begin{figure*}
     \centering
      \includegraphics[width=.75\textwidth]{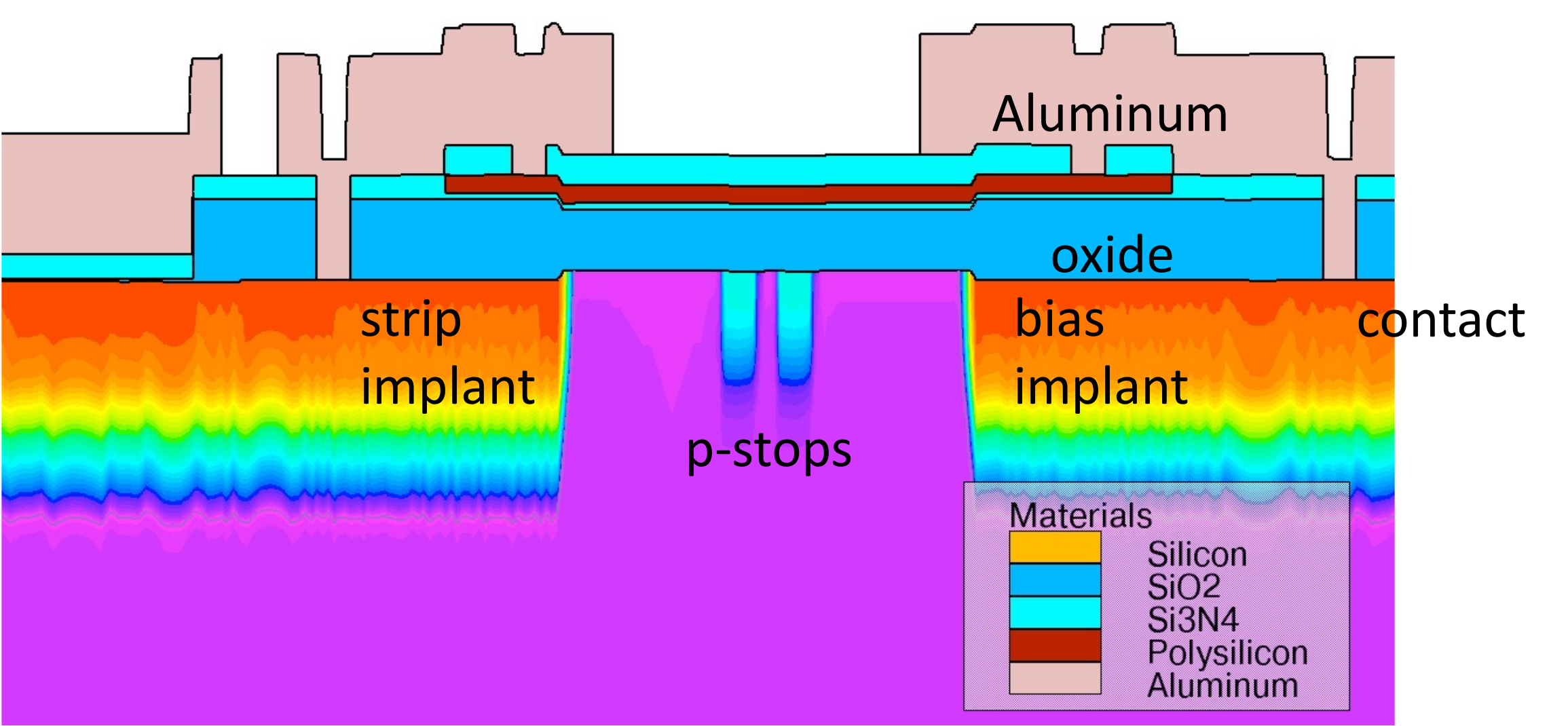}
    \caption{Results of the process simulation of the strip end region of an  
    AC coupled detector including implants, AC coupling layers, and polysilicon
    resistors.} 
\label{AC_8}
\end{figure*}

Commercial TCAD packages are capable of full simulations of device fabrication, including 
epitaxy, implantation, annealing, deposition and oxidation.  The accuracy and detail provided by these 
simulations can be invaluable in the 
development of new sensor technologies or in understanding the behaviour of existing devices. 
For devices such as LGADs, an accurate model of the implant doping density and profile is crucial to understanding and predicting performance. This is possible with process simulation as the simulation includes detailed models of the implantation and annealing 
 including crystal channeling, implantation damage, and dopant activation. 
The process  model can also provide a link between commercial foundries and instrumentation developers. Figure~\ref{AC_8} shows the result of the process simulation
for 8-inch AC-coupled wafer development. The wafers were fabricated by NHanced/Novati 
with process flow input from LBNL and SLAC and the Silvaco process simulation from Fermilab \cite{Alyari:2020pet}.

\subsection{Hamburg Model Simulation of Silicon Radiation Damage}\label{HamburgModel}
\subsubsection{Modeling Silicon Leakage Current: the Hamburg Model}\label{LC_Hamburg}


One of the best-characterized methods for monitoring silicon radiation damage is based on measuring the sensor leakage current.  The most widely used-model for predicting the leakage current is the Hamburg Model~\cite{moll-thesis}. A linear relationship between particle fluence and leakage current applies to silicon sensors. The effective fluence, $\Phi_{\mathrm{eq}}$, is the number of $1$-MeV neutrons applied to a sensor of surface area $1$ cm$^2$ that cause damage equivalent to that of all particles that went through the sensor. The linear relationship between effective fluence and $\Delta I$, the change in leakage current at fluence $\Phi_{\mathrm{eq}}$ relative to the value before irradiation, is given as
\begin{equation}
\Delta I = \alpha\cdot\Phi_{\mathrm{eq}}\cdot V\,,
\end{equation}
where $V$ is the depleted volume of the silicon sensor and $\alpha$ is the current-related damage coefficient~\cite{Lindstrom:2002gb} which is a function of annealing time and temperature. The observed leakage current and predictions are typically normalized to the same temperature for direct comparison. The normalization is made using the known leakage current dependence on temperature and the effective silicon band gap energy, $E_\mathrm{eff}$~\cite{Chilingarov_2013}.
The change in the leakage current is predicted using the Hamburg Model by inputting the complex radiation fields in the detector in fluence simulated by packages such as \textsc{Fluka}~\cite{Baranov:2005ewa, Battistoni:2007zzb} and \textsc{Geant}4~\cite{GEANT4:2002zbu, Allison:2006ve, Allison:2016lfl}, as given by the following equation:
\begin{equation}
\Delta I = \left(\Phi_{\mathrm{eq}}/L_{\mathrm{int}}\right)\cdot V\cdot\sum_{i=1}^nL_{\mathrm{int},i}\cdot\left[\alpha_I\exp\left(-\sum_{j=i}^n\frac{t_j}{\tau(T_j)}\right)+\alpha^*_0-\beta\log\left(\sum_{j=i}^n\frac{\Theta(T_j)\cdot t_j}{t_0}\right)\right],
\label{eq:hm}
\end{equation}
where $L_{\mathrm{int}}$ is the integrated luminosity, $t_i$ is the time, and $T_i$ is the temperature in period $i$. The first sum is over all time periods and the two sums inside the exponential and logarithmic functions are over the time between the irradiation in time period $i$ and the time of the measurement. The other symbols in Eq.~\ref{eq:hm} are $t_0=1$ min, $V=$~depleted sensor volume (in $\mathrm{cm}^3$), $\alpha_I=(1.23\pm 0.06)\times 10^{-17}$ A/cm, $\tau$ follows an Arrhenius equation $\tau^{-1}=(1.2^{+5.3}_{-1.0})\times 10^{13}~\mathrm{s}^{-1}\times\exp^{(-1.11\pm0.05)/k_\mathrm{B}T}$ (where the units of $k_\mathrm{B}T$ are eV), $\alpha^*_0=7.07\times 10^{-17}$A/cm, and $\beta=(3.29\pm 0.18)\times 10^{-18}$ A/cm. The temperature scaling function $\Theta(T)$ is defined by
\begin{equation}
\Theta(T)=\exp\left[\frac{-E_\mathrm{eff}}{k_\mathrm{B}}\left(\frac{1}{T}-\frac{1}{T_\mathrm{R}}\right)\right],
\end{equation}
where $E_\mathrm{eff}=1.21$ eV is used for the effective silicon band gap energy, $k_\mathrm{B}$ is the Boltzmann constant, and $T_\mathrm{R}=21^\mathrm{o}$C. Once the simulation is complete, a temperature correction to the full simulation is made so that it may be compared to the leakage current data.

\begin{figure}
\centering
\includegraphics[scale=0.5]{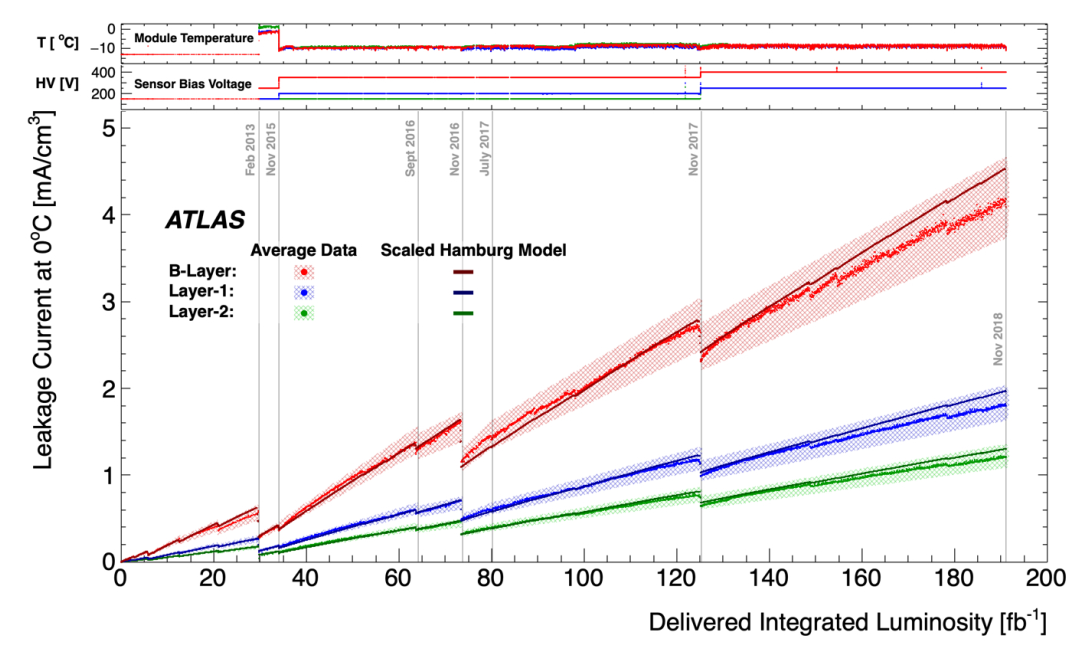}
\caption{Average measured leakage current of a representative sample of modules on the B-Layer, Layer-1 and Layer-2 over the full period of operation. The scaled prediction from the Hamburg Model is also shown. The bands include uncertainties on the measurement~\cite{ATLAS:2021gld}.}
\label{fig:leakage_current_barrel_layers}
\end{figure}

The leakage current predictions~\cite{ATLAS:2021gld} for ATLAS outer barrel layers - B-Layer, Layer-1, and Layer-2 are shown in Figure~\ref{fig:leakage_current_barrel_layers}. The average of measured leakage currents of all layers are shown for the comparison. It is clear from the plot that the Hamburg Model correctly describes the leakage current evolution of the ATLAS Pixel detector, after scaling to the data.

\subsubsection{Modeling Depletion Voltage and Charge Collection Efficiency}\label{Vdep_CCE_HM}
\begin{figure}
\centering
\includegraphics[scale=0.2]{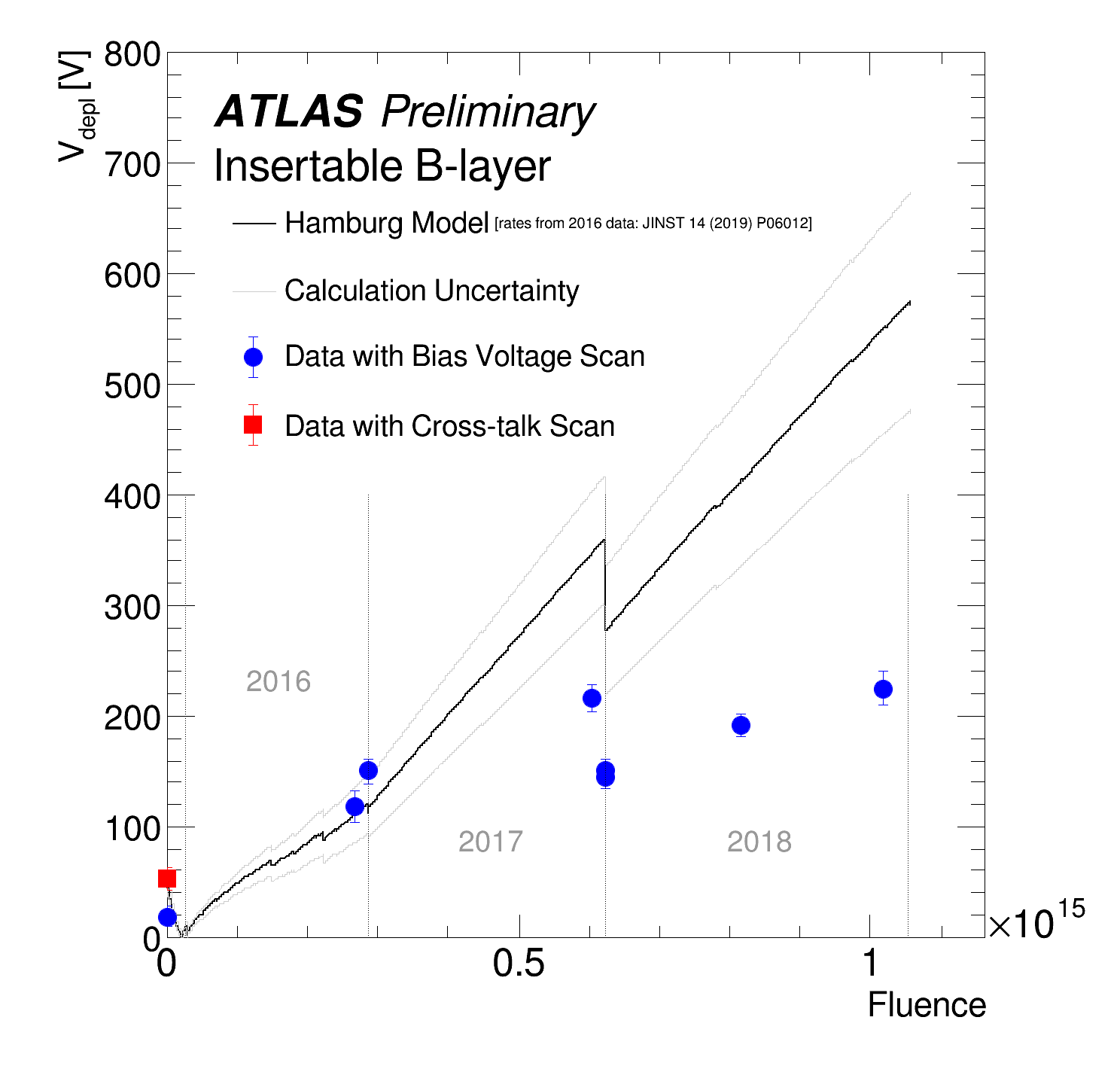}
\caption{Predicted~\cite{Dawson:2764325} 
depletion voltage of the IBL according to the Hamburg model as a function of time from installation until the end of Run 2. Circular and square points are measured depletion voltages.}
\label{fig:IBL_V_dep}
\end{figure}

The full depletion voltage $V_\mathrm{dep}$ of a sensor depends linearly on the doping concentration $N_\mathrm{eff}$ and the square of the thickness $d$ of the detector (wafer thickness for a planar sensor or electrode-to-electrode distance for a 3D sensor), and can be written as
\begin{equation}
V_\mathrm{{dep}}=\frac{\mid N_\mathrm{eff}\mid d^2q}{2\epsilon\epsilon_0},
\end{equation}
where $q$ is the charge of the electron, $\epsilon$ is the dielectric constant, and $\epsilon_0$ is the vacuum permittivity. The doping concentration is affected by the fluence, and the change in doping concentration $\Delta N_\mathrm{eff}$ can be expressed as
\begin{equation}
\Delta N_\mathrm{eff}= N_\mathrm{A} + N_\mathrm{Y} + N_\mathrm{C},
\label{Eq:doping_conc}
\end{equation}
where $N_\mathrm{A}$, $N_\mathrm{Y}$, and $N_\mathrm{C}$ are the changes in the doping concentration due to beneficial annealing, reverse annealing, and stable damage respectively. The Hamburg Model\footnote{This model is not directly related to the Hamburg Model for leakage current, even though they share the same name.} predicts the depletion voltage of a sensor when it is exposed to radiation causing non-ionizing energy loss (NIEL) for a certain duration, by using the change in the doping concentration described by Eq.~\ref{Eq:doping_conc}. However, the prediction requires a set of parameters called introduction rates which depend on the material of the detector and  the type of radiation the detector receives. The existing predictions of depletion voltage use introduction rates calculated using data recorded at relatively lower fluences~\cite{moll-thesis} and are not suitable for predictions of depletion voltage of detectors designed for future conditions. As an illustration of this limitation, a comparison~\cite{Dawson:2764325} 
of depletion voltage prediction to measured depletion voltage of the ATLAS Insertable B-Layer is shown in Figure~\ref{fig:IBL_V_dep}. 
As of now,
the full ATLAS Pixel sensor depletion voltages recorded in Run 2 are accessible, so the Hamburg Model can be used in a fit to the data to extract these parameters; this information can be projected to higher fluences, and used for developments of future particle physics detectors.

An important consideration is the loss of charge collection efficiency (CCE). CCE loss due to radiation becomes the leading failure mode at future colliders. In the current implementation of the Hamburg Model, prediction of charge collection is not included.  Predicting the charge collection efficiency loss of subdetectors, based on received fluence and annealing time, is essential to ensure an efficiently functioning detector.
\subsection{RASER -- SiC detector simulation}\label{RASER} 
To explore the potential capability of wide band gap semiconductor especially silicon carbide (SiC) to be used in high energy physics experiments, a fast simulation software -- RASER (RAdiation SEmiconductoR)\footnote{https://pypi.org/project/raser} has been developed as a python module. 
In RASER, the electric field and weighting potential are calculated by FEniCS~\cite{FEniCS} -- an open-source computing platform for solving partial differential equations. The simulation of electrons from $^{90}$Sr and the deposited energy in SiC are carried out by \textsc{Geant}4. The current induced by electron--hole ($e$--$h$) pairs moving is calculated by Shockley-Ramo's theorem~\cite{shockley,ramo}. The readout electronics used a simplified current amplifier as in~\cite{weightfield2}.  

\begin{figure*}
     \centering
      \includegraphics[width=.4\textwidth]{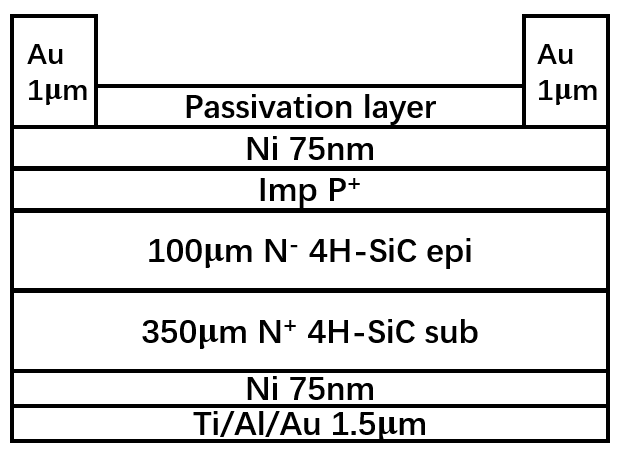}
    \caption{Cross section of $\rm 5~mm\times5~mm$ planar 4H-SiC detector designed by NJU. It has $\rm 100~\upmu m$ 4H-SiC active epitaxy layer and a $\rm 350~\upmu m$ substrate. Figure from \cite{sicpin}. 
    } 
\label{fig:njusic}
\end{figure*}

\begin{figure*}
     \centering
      \includegraphics[width=.6\textwidth]{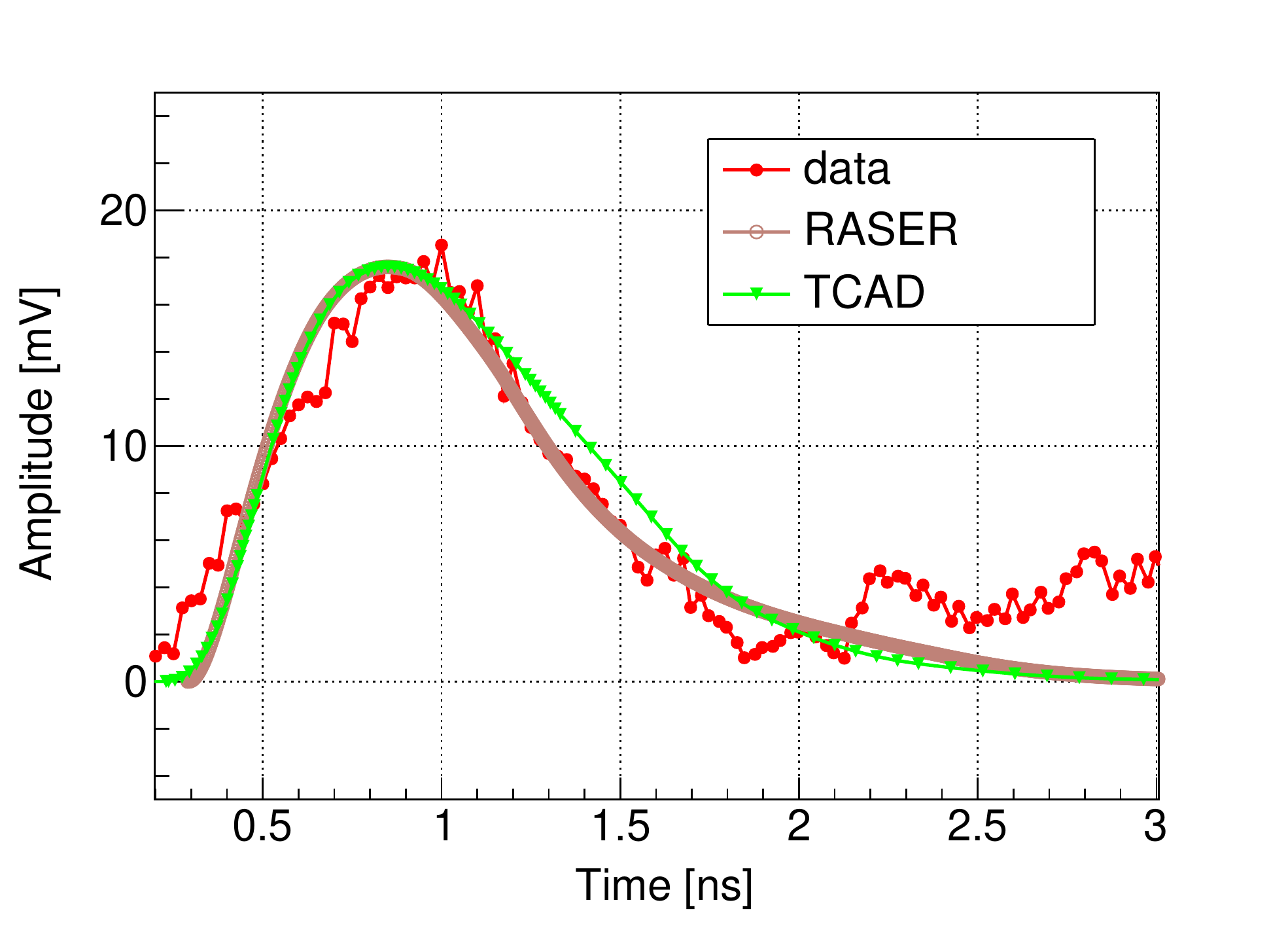}
    \caption{Waveform comparison: measured data (red) and RASER simulation (green). Figure from \cite{sicpin}.
    } 
\label{fig:compare_sim_mea}
\end{figure*}

RASER has been validated by comparing measurement and simulation of time-resolution of planar 4H-SiC detector designed by Nanjing University (NJU) in China. The cross section of the SiC detector is shown in Figure~\ref{fig:njusic}. The size of the detector is 5 mm$~\times$ 5 mm, and the upper and lower electrodes are ohmic contacts. The detector has a 100 µm high resistance active 4H-SiC epitaxial layer and a 350 µm substrate. The comparison between wave-forms obtained from measurement, RASER, and TCAD simulation shows good consistency as in Figure~\ref{fig:compare_sim_mea}. Using the same parameters, the time resolution of measurement and simulation are (\mbox{94 $\pm$ 1})~ps and (73 $\pm$ 1)~ps, respectively~\cite{sicpin}. 

\begin{figure*}

\subfloat[ ]{
\includegraphics[width=.25\textwidth]{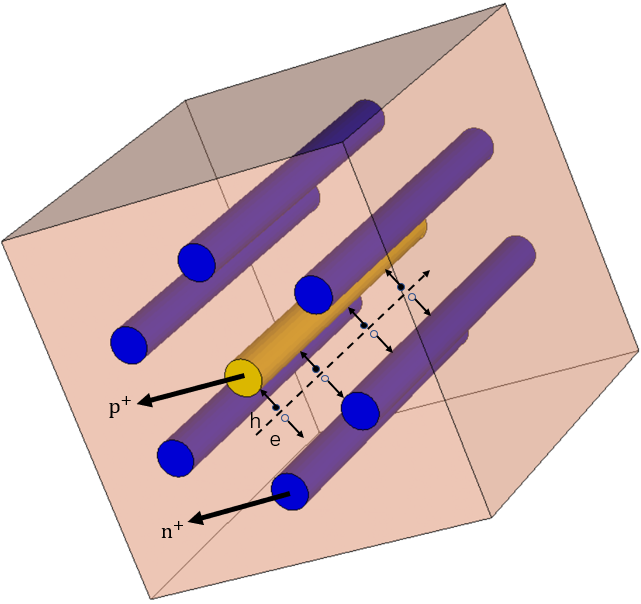}\label{fig:3dsicA}
}
\subfloat[ ]{
\includegraphics[width=.25\textwidth]{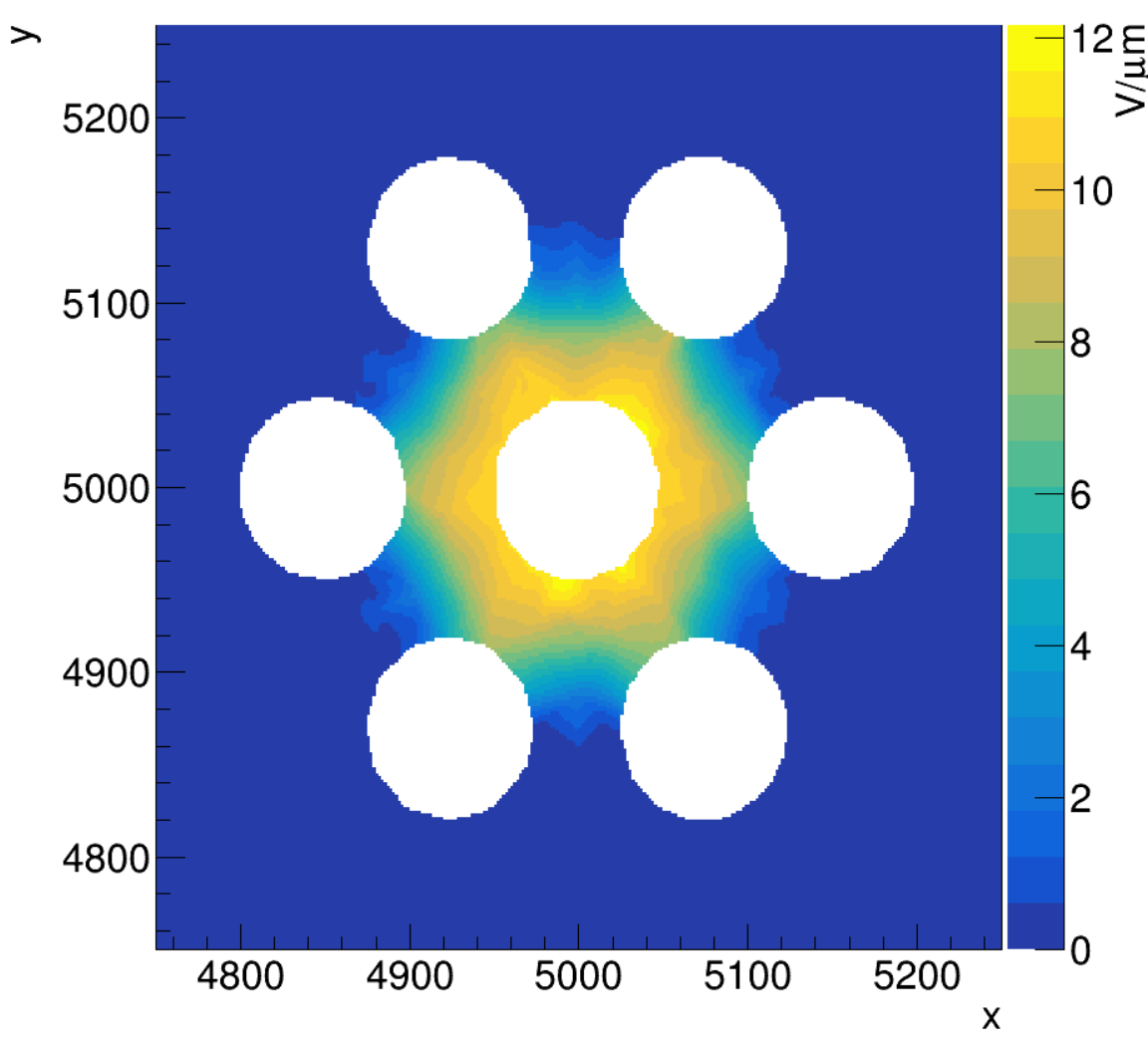}\label{fig:3dsicB}
}
\subfloat[ ]{
\includegraphics[width=.40\textwidth]{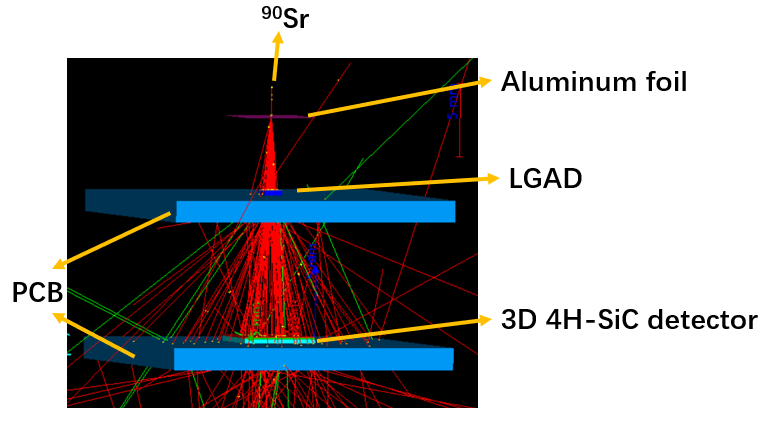}\label{fig:3dsicC} 
}
\caption{ (\textbf{a}) 3D 4H-SiC detector schematic diagram. (\textbf{b}) The electric field distribution of x--y cross section for the 3D 4H-SiC detector with 500 V bias voltage simulated with RASER. (\textbf{c}) \textsc{Geant}4 simulated two electrons passing through aluminum foil, the LGAD, a printed circuit board (PCB), and the~3D 4H-SiC detector. The tracks of electrons and the deposited energy with a step less than 1~µm were recorded. Figure from \cite{3D-SiC}. 
}

\label{fig:3dsic}
\end{figure*}

RASER has also been used to predict the timing performance of 3D-SiC with a potential of more radiation hard that could operate at room temperature. The first structure of 3D 4H-SiC detector prototype studied is shown in Figure~\ref{fig:3dsicA}, where the $p^+$ electrode column that penetrates the entire high-resistance substrate perpendicular to the detector surface is surrounded by six $n^+$ electrode columns. The simulated size of the 3D 4H-SiC detector is 1~cm~$\times$~1~cm, and the thickness is 350~µm, and~the effective concentration of n-type substrate is set to 1~$\times$~10$^{13}$~cm\textsuperscript{$-$3}. The radius of electrode column is 50~µm, and~the column spacing, which is the distance between the center of $n^+$ and $p^+$ electrode columns is 150~µm. The simulated electric field is shown in Figure~\ref{fig:3dsicB}. The 3D-SiC detector is placed with a 34 ps LGAD under the $^{90}$Sr source where the emitted two electrons are simulated by \textsc{Geant}4 as shown in Figure~\ref{fig:3dsicC}. 

\begin{figure*}

\subfloat[ ]{
\includegraphics[width=.45\textwidth]{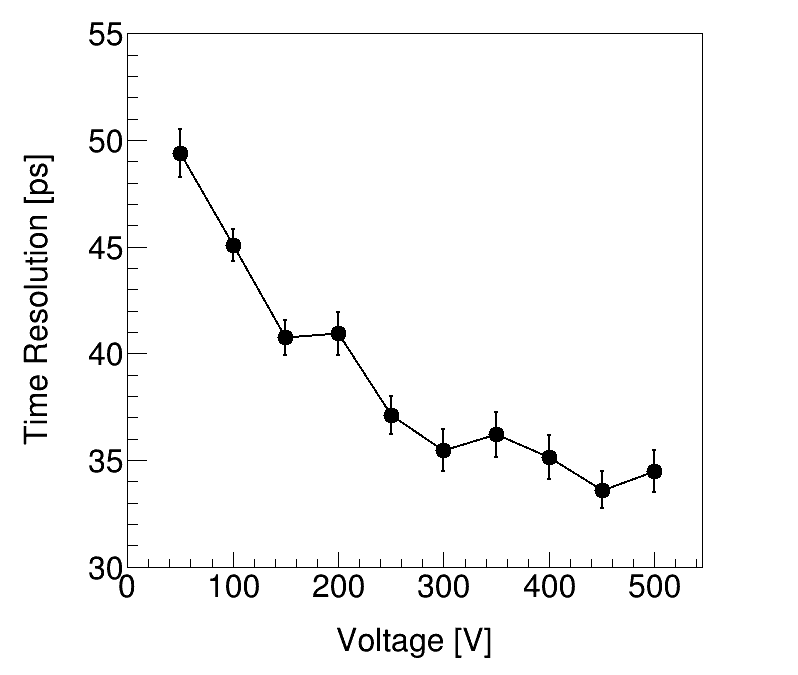}\label{fig:3dsictimeA}
}
\subfloat[ ]{
\includegraphics[width=.45\textwidth]{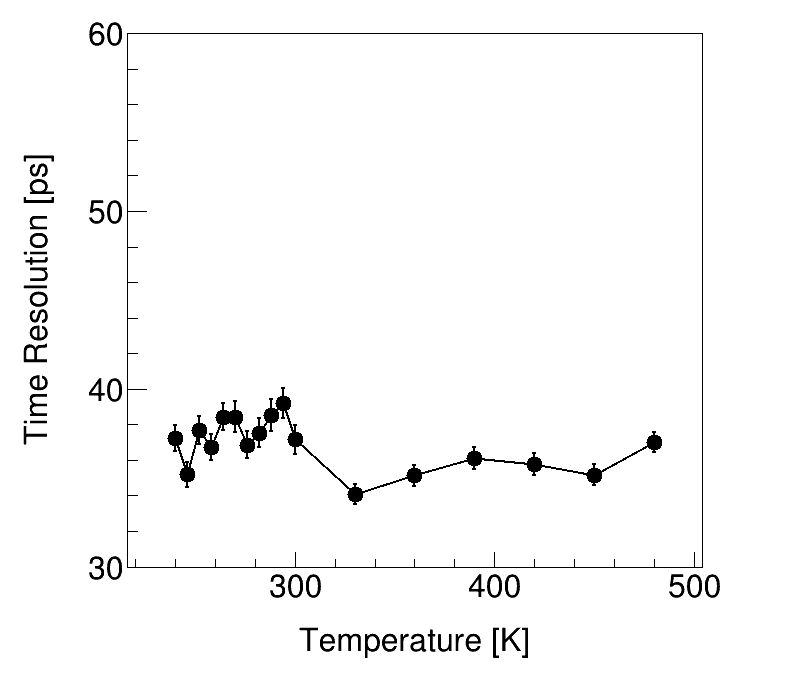}\label{fig:3dsictimeB}
}
\caption{ (\textbf{a}) Time resolution versus voltage. (\textbf{b}) Time resolution versus temperature. Figure from \cite{3D-SiC}.}

\label{fig:3dsictime}
\end{figure*}

The simulated time resolution reached around 35 ps with bias voltage of 300 V (Figure~\ref{fig:3dsictimeA}) while remaining constant with temperature even with values up to +127$^{\circ}$ C (Figure~\ref{fig:3dsictimeB}). Simulation of 5 electrodes configuration (3D-4H-SiC-5E) reveals up to 25 ps time resolution can be reached~\cite{3D-SiC}. 

Future development of RASER will include the LGAD-SiC as well as the pixel-timing performance for SiC devices. 

\subsection{Additional Capabilities}\label{AddCapab}
In addition to core semiconductor modelling commercial packages often have additional functionality for specialized applications. These include optoelectronic simulation,
lattice heating simulators, single event effects, VCSEL and LED simulation, quantum effects simulation and a number of other options.  These are typically individually licensed.

\begin{figure*}
     \centering
      \subfloat[ ]{
      \includegraphics[width=.5\textwidth]{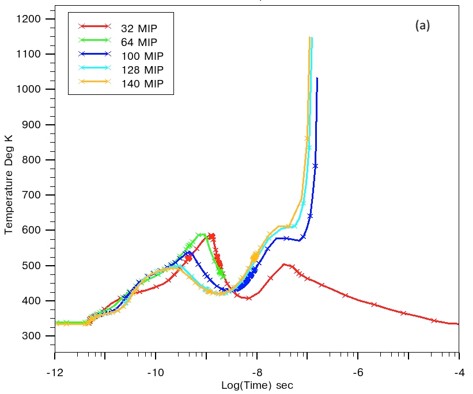}}
      \subfloat[ ]{
      \includegraphics[width=.5\textwidth]{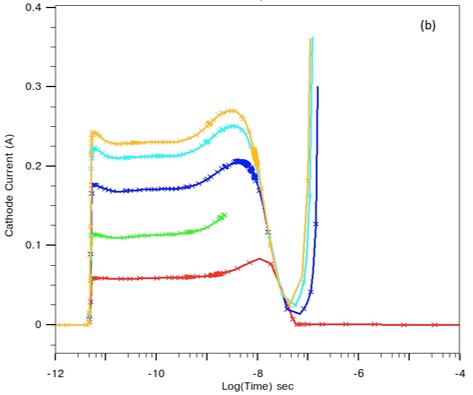}}
    \caption{Transient simulation of junction temperature (a) and cathode current (b) for a 50-$\upmu$m-thick diode operated at 625 V exposed to heavily ionizing particles between 32 and 140 MIPs. The threshold for burnout is between 64 and 100 MIPs. There is about a 100 microsecond delay between impact and the destructive heating of the sensor.} 
\label{SEB}
\end{figure*}

An example of such a simulation in HEP is single event burnout. Test beam studies of 
irradiated LGADs showed a high rate of beam-induced breakdown resulting in physical 
destruction of the device. The Silvaco Giga lattice heating toll was used in 
combination with avalanche models to study the systematics  of these effects, including 
linear energy transfer threshold and the effect of the field strength on breakdown 
margin \cite{2010270}. Figure~\ref{SEB} shows transient curves of temperature and cathode current for a 
50-$\upmu$m-thick diode operated at 625 V struck by particles with 
ionization between 32 and 140 MIPs. 

\subsection{Testbeam}
\label{sec:testbeam}
Sensor simulations from commercial packages like TCAD 
are less often used for Monte Carlo simulations of drift and diffusion of carriers or, for example, simulations of beam tests at facilities with charged particle beams. For such purposes, specialized Monte Carlo software is used. These Monte Carlo simulation packages can use sensor simulations results such as the sensor electric field as input; see for example \cite{allpix2withTCAD}.

One example of a software package for drift and diffusion modeling of carriers in silicon sensors is \texttt{PixelAV} \cite{Chiochia2005, Swartz2006212, Swartz:687440}, originally created for interpretation of data taken with both unirradiated and irradiated sensors at pion test beams. This package has accurate models of charge deposition and transportation and can simulate charge drift under magnetic fields. It also models radiation damage with the trapping of charges. It uses input electric fields from the TCAD 
simulation software. \texttt{PixelAV} is now used as standalone software in the CMS template-based hit reconstruction software that includes radiation damage simulation \cite{Swartz:2007zz, Chiochia:2008yga, CMS:2014pgm}. In the templates created with \texttt{PixelAV} provide both corrections to the hit position and the cluster charge. 

Packages similar to \texttt{PixelAV} are \texttt{Allpix} \cite{allpix} 
and \texttt{KDetSim}. 
\texttt{KDetSim} is based on the CERN \texttt{ROOT} software package. It is fast, allows for simulation of larger volumes, and allows for iterative approaches. All packages are free and open source.

\begin{figure*}
     \centering
      \includegraphics[width=.7\textwidth]{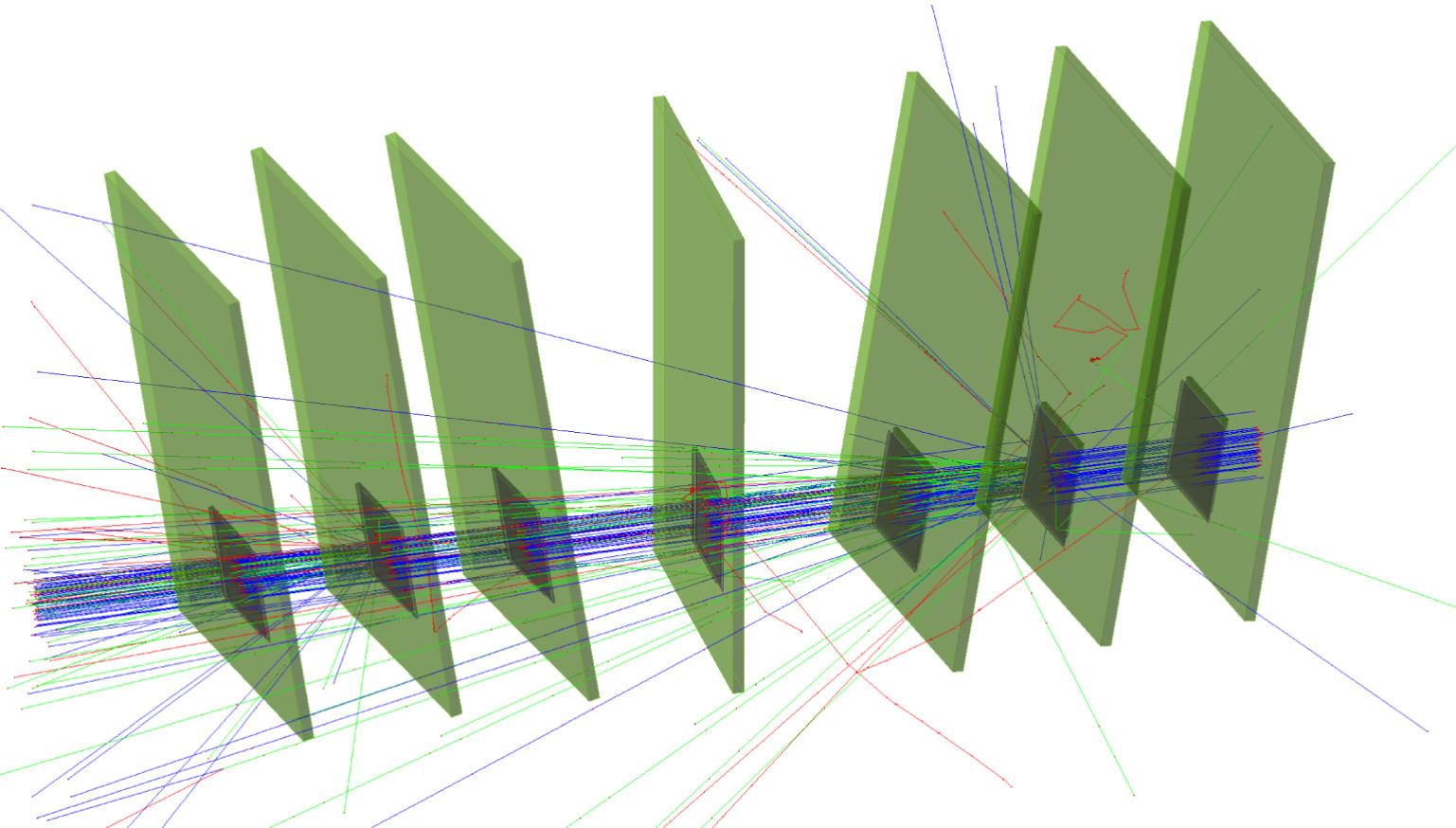}
    \caption{Visualization of a simulated detector telescope setup with 7 planes. in \texttt{Allpix$^2$}. The beam is incident from the right, and the colored lines are the primary and secondary particles propagated through the setup by the simulation. Figure from \cite{allpix2}.
    } 
\label{fig:allpix2_visualization}
\end{figure*}

\begin{figure*}
     \centering
     \subfloat[ ]{
      \includegraphics[width=.42\textwidth]{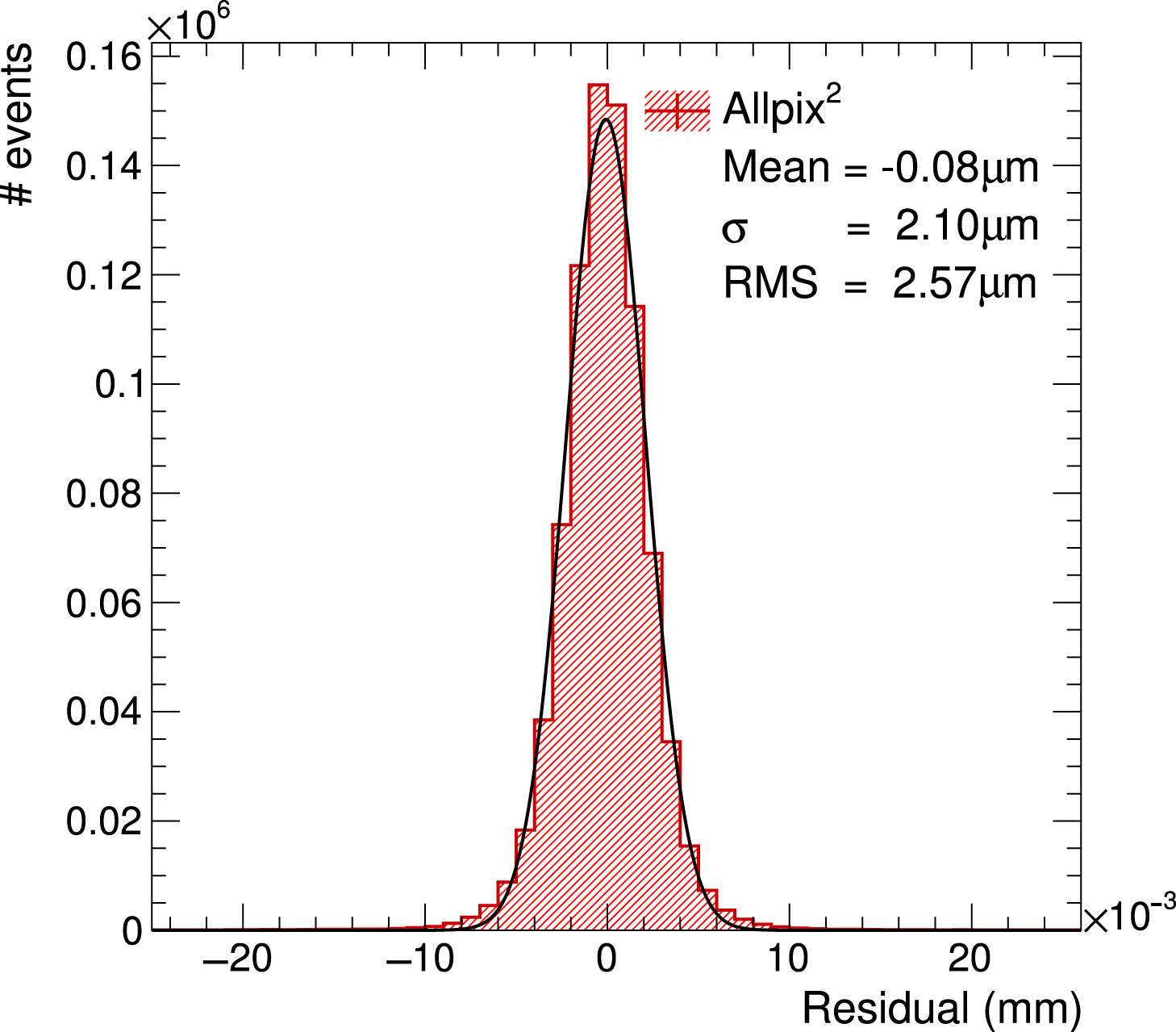}}\hspace{1 mm}
      \subfloat[ ]{
            \includegraphics[width=.4\textwidth]{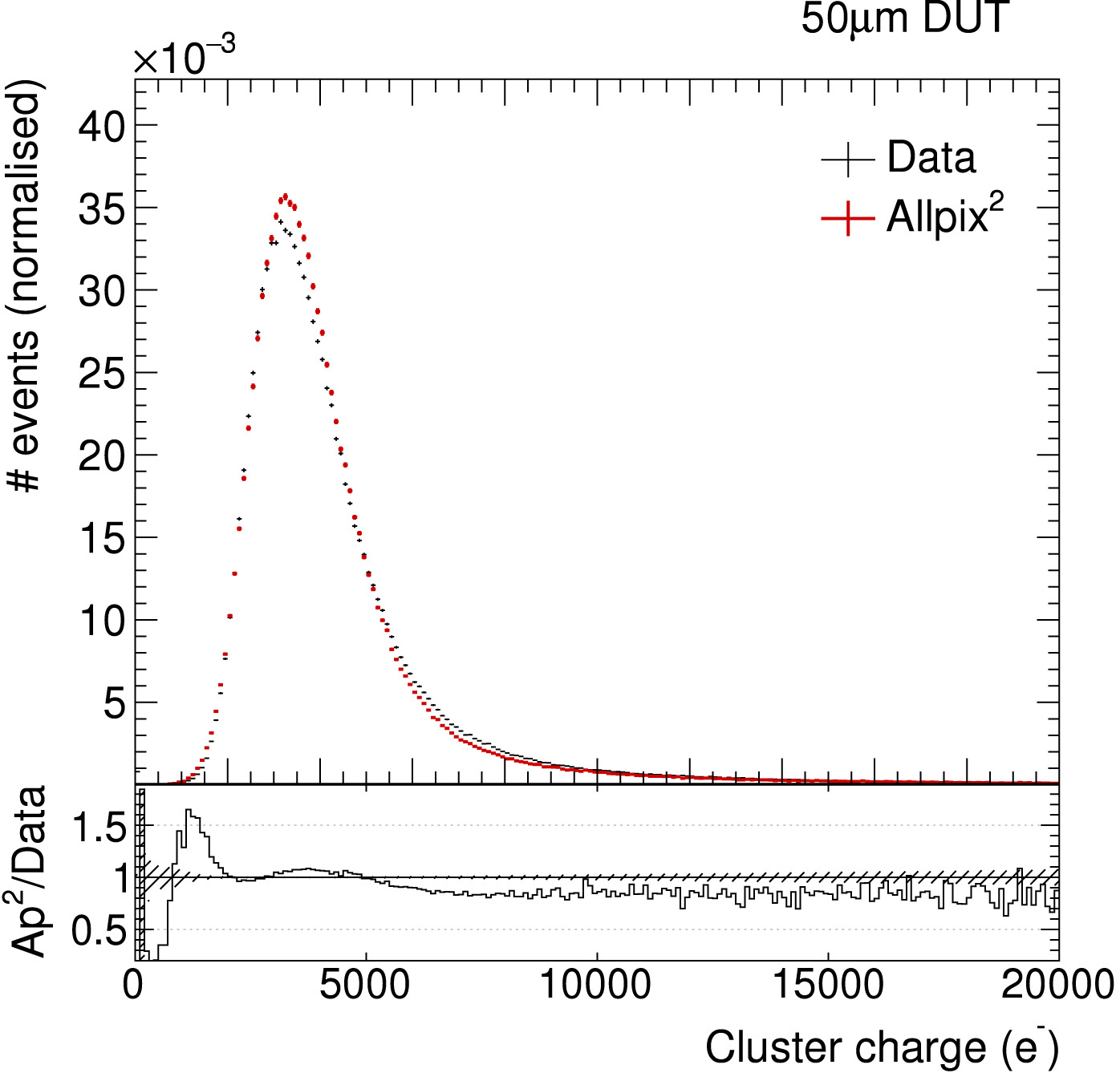}}
    \caption{(a) Residuals and (b) cluster charge distribution simulated with the \texttt{Allpix$^2$} software package compared to data. The cluster charge is for a 50 $\upmu$m device under test (DUT). The residuals are the difference between the reconstructed telescope track position in the DUT and the true hit position of the simulated particle. Figures from \cite{allpix2}.
    } 
\label{fig:allpix2_residuals_clustercharge}
\end{figure*}

Another free and open source package that includes experimental setups such as beam telescopes and their material effects like multiple scattering and nuclear interactions based on \texttt{Geant4} is \texttt{Allpix$^2$} \cite{allpix2}. This includes energy deposition based on \texttt{Geant4}, drift and diffusion for charge propagation, and models charge digitization. A visualization of a setup provided by the software package is shown in Fig. \ref{fig:allpix2_visualization}.
The \texttt{Allpix$^2$} framework can also provide observables for comparison with test beam data such as particle clusters, tracks, residuals, and cluster charge. An example is shown in Fig. \ref{fig:allpix2_residuals_clustercharge}.
The packages now also include different charge carrier mobility models and the possibility to load doping profiles for doping-dependent charge carrier lifetime calculations \cite{spannagel2021allpix2}, and foresees to include trapping of charges.

\subsection{Full detector systems}\label{FullDetSys}


Within large HEP collaborations, such as those of the 
LHC-experiments, silicon detector simulation is typically performed using dedicated, proprietary\footnote{This is changing - many collaborations now make their source code public.} software written within the experiment's software framework (for example Athena\footnote{https://gitlab.cern.ch/atlas/athena}, Gauss\footnote{https://gitlab.cern.ch/lhcb/Gauss}, and CMSSW\footnote{https://github.com/cms-sw/cmssw} for ATLAS, LHCb and CMS, respectively). Even within a single framework, multiple different approaches may exist which have been tailored for a specific use-cases.

Detailed detector performance studies typically need the most detailed approach available, which comes at the cost of significant highly processing times. Such studies can generally be performed to a sufficient level of precision by using smaller samples sizes and/or a limited set of physics processes. This is not the case for the production of Monte Carlo samples for use in data analyses, where large samples covering a wide range of signal and background processes are needed.

Using fully-detailed approaches for production of physics analysis Monte Carlo samples can therefore sometimes become prohibitive, and this will increasingly become the case for High-Luminosity LHC. Faster and more approximate approaches therefore also need to be made available to be used in the production of such samples, potentially with different approaches being used for signal processes to those used for backgrounds, or the minimum bias collision events used to represent the large number of additional `pile-up' interactions per bunch crossing which accompany the signal process. In certain circumstances, different levels of detail may be considered at different positions within the detector, with a more accurate modeling applied for crucial measurements close to the particle interaction point, or in areas of high particle flux.

The faster, more approximate methods will typically rely on a parameterization or templates derived from the fully-detailed approach, stand-alone simulations, or from data. Appropriate and robust procedures for making comparisons of key observables between the different approaches and also crucially between experiment-specific and stand-alone software are therefore highly beneficial for this process.

One aspect of silicon detector simulation which is becoming increasingly important is the modeling of radiation damage. 
Different approaches to this topic include\footnote{See also \href{https://indico.cern.ch/event/754063/contributions/3222832/attachments/1758817/2852814/RD50TalkNovember2018.pdf}{this talk} for an overview.}:

\begin{itemize}
    \item A modified digitization approach has been employed by ATLAS, in which specific adaptations are made within the code responsible for modeling charge generation and transport within the sensor, in order to account for the effects of radiation damage on the generated charges prior to their readout~\cite{RadDamageDigi_2019}.
    \item CMS takes a different approach, in which \emph{cluster templates} are derived based on comparing the properties of charge clusters produced in independent simulations with and without radiation damage effects. These templates are parameterized according to pertinent factors such as particle incident angle, and position within the detector, in order that appropriate efficiency corrections can be applied to channels within the clusters produced without explicit radiation damage modeling applied.
    \item LHCb uses an even faster approach whereby fundamental silicon parameters (such as the diffusion) are tuned to data to reproduce the effective consequences of radiation damage.  For example, a multiplicative factor is used to reduce the predicated collected charge and the diffusion length parameter is increased to match data.  Currently, these parameters are updated annually. 
\end{itemize}

\section{Challenges and Needs}\label{CandN}
\subsection{Proprietary software and device properties}\label{PropSoftware}
One of the main systematic uncertainties in device simulation is the doping profile of a sensor. This can both be the doping profile of a sensitive layer such as an epitaxial layer in a monolithic sensor, or the doping profile of an implant  \cite{munker_thesis}. These properties of a sensor are often proprietary and are adjusted in simulations through tuning the simulation to measurements. Also, to understand the effect of the uncertainty, doping profiles of a sensor varied in both electrostatic and transient simulations.
    
In addition, device simulation software like TCAD 
is proprietary, leaving the possibility for use up to whether an institute can afford a license and no possibility for writing extensions. This also creates a dependence on only one correct installation of the software per institute, as personal installation is not possible. The \texttt{Weightfield2} \cite{weightfield2} package provides an open source alternative to TCAD 
that requires much less computing power but is also not a replacement for the quite accurate device simulation in TCAD. 

\subsection{Computational requirements of TCAD-simulations}\label{ComputTCAD}
Two-dimensional electrostatic and transient simulations with software like TCAD 
can often provide a good approximation of the electric and weighting fields in a three-dimensional sensor. A sensor design is translated into a mesh of points, where computations are done for each point in the mesh. To get a more accurate picture of a sensor, three-dimensional simulations are also possible. These results, however, in a much larger amount of mesh points for which each step of the finite element simulation is stored. To optimize the number of mesh points, one can optimize the position of a representation mesh of a sensor using boundary conditions where edge effects are avoided.

\subsection{Combined TCAD 
and transient simulation tools}\label{TCAD_TCT}
As described in \ref{sec:testbeam}, Monte Carlo simulation tools such as \texttt{Allpix$^2$} \cite{allpix2}, \texttt{KDetSim} \cite{kdetsim1}, and \texttt{TCode} \cite{Loi2020} can be combined with electrostatic and transient simulations from device simulation tools like TCAD. 
\texttt{Allpix$^2$} has an approximation of a linear electric field, but this is in the case for monolithic sensors with small collection electrodes, for example, not enough to describe the complex electric fields resulting from the sensor design. The sensor design needs to be implemented precisely, and knowledge from planar sensors cannot be transferred. A two-dimensional or three-dimensional electric field as well as weighting field of a sensor can be fed into the \texttt{Allpix$^2$} software but here care needs to be taken for it to be in the correct format. The \texttt{Allpix$^2$} package includes a converter for a certain mesh file format produced with TCAD 
simulations and plans for an interface with \texttt{Weightfield2} \cite{spannagel2021allpix2}, but an all-in-one package does not yet exist: there is still a need to set up simulation chains.

\subsection{Uncertainties in TCAD models}\label{uncertTCAD}
Uncertainties in TCAD 
models, both systematic and statistical, are not always used and mentioned consistently, nor can they always be estimated well. As described above, doping levels of silicon sensors can influence electric fields, the depletion region, and signal response. These are, however, very often proprietary and not known to the author of the simulation. Temperature can influence leakage currents, and is also under certain circumstances not known. Different charge carrier lifetimes can also influence leakage currents, as well as signal response. The position of the grid on which the finite element simulation is computed can, if not fine enough, yield inaccurate results. Finally, using two or three dimensions affect the accuracy of the simulation but also the computing time.

\subsection{Unified radiation damage (TCAD) and annealing}\label{unifiedTCAD}

The detailed mix of particle types and energies can have a 
strong impact on the level of macroscopic radiation damage  effects and even more on the damage at microscopic level, in particular on the electric field profile inside the sensor bulk~\cite{Kramberger_2014,MikuzVertex2017}. 
So, even if a unified TCAD radiation damage model would be 
desirable, still different models are needed, depending on the type of irradiation. 
For example, from the experience of silicon detectors at hadron colliders (LHC, HL-LHC) it would be good to derive TCAD radiation damage models knowing in detail the mix of particle types and energies that had crossed the detectors during the data taking periods. This knowledge could be then used for the future experiments ad hadronic colliders (FCC-hh), but once the expected cocktail of irradiation has been carefully simulated.

At the moment no TCAD radiation damage model includes 
a parameterization for annealing effects. Often the 
TCAD radiation damage models have been based on data 
from irradiated sensor that underwent after irradiation 
some annealing process before being measured~\cite{Chiochia2005,FOLKESTAD201794,schwandt2019new}.  The typical annealing process for irradiated test 
structures is the so-called ``optimal annealing'' (80 minutes at 60$^{\circ}$~C), which assures achieving the 
smallest depletion voltage for a fixed level of irradiation. In installed and running detectors it is difficult to have 
a precise annealing scheme: if it is true that in LHC experiments silicon trackers are operated cold ($< 0^{\circ}$~C), the temperature during maintenance is more difficult to be controlled, hence some unwanted annealing can take place. Simulating this effect in TCAD turns out to be difficult as it could be already being taken into account by the original model. 

For what concerns experiments at hadronic colliders, in the simulation of radiation damage to ATLAS Pixels~\cite{RadDamageDigi_2019} an effective approach has been adopted, which consisted in matching the predicted
 annealing effect~\cite{moll-thesis} on the effective doping concentration $N_{eff}$ with the concentration of 
 simulated deep-defects. The effect turned out not to be so large thanks to the fact that ATLAS Pixels detector was kept cold most of the time during its operation.  This may not always be the case and for significant annealing effects, more sophisticated approaches are required. 
 \subsection{Improving the Hamburg Model Simulation}\label{Improve_HM}
The Hamburg Model code is written in C++ and utilizes tools in ROOT. Both leakage current and depletion voltage predictions using the Hamburg Model require as inputs the temperature of the sensor, fluence, and duration of the fluence application on the detector. The code iterates through all temporal measurement points to predict the leakage current and depletion voltage at a future time. The time taken for the iteration depends on the total duration of the fluence application which is the operational period of the detector. This can lead to unsupportable run-time for long-term predictions. However, the code can be accelerated. The methods for accelerating the code are in development. 

Addition of charge collection efficiency prediction routines is also proposed. Furthermore, geometric generalization that allows modeling of alternative sensor geometries (such as 3D versus planar) is needed. A multi-threaded implementation of the code will significantly improve the performance of the code in iterating through all fluence points.
  \subsection{Damage Models}
  One central uncertainty that obfuscates the translation between predicted fluence and sensor damage is the precision of sensor damage factors. These factors describe how much damage is introduced by a particle of a given energy and given type. Many testbeam measurements have been made to constrain these parameters, but most measurements are not produced with uncertainties and many factors at energy values away from beam energies are based on simulations or other extrapolations. The RD50 Collaboration maintains a database of damage factors\footnote{https://rd50.web.cern.ch under NIEL.}, but a dedicated measurement program is needed to modernize and improve them. See Ref.~\cite{ATLAS:2021gld} for additional challenges associated with sensor modeling and with current unexplained systematic differences between data and predictions.

\subsection{Challenges of TCAD radiation-defect modeling for future colliders}\label{TCADfuture}
For more complete parameterizations of radiation-induced properties like CCE($x$), test beam measured charge collection data of varying pitches of both strip and pixel detectors as well as higher number of fluence points would be required. 
Future efforts of the defect model developments should also include further calibration with the indirect electric field information from the measured edge-Transient Current Technique (TCT) data that enables the tuning of the simulated $E$($z$) profile with increased precision. 

The main tuning effort in the TCAD defect models discussed in Section~\ref{TCAD} has typically been focused to a fixed polarity and 
geometry device, leading to compromised accuracy of the simulation results when 
these are varied. 
A more general expression of device parameters would provide a starting point towards unified defect models. Furthermore, effort should be put to tune the standard parameter sets of the Sentaurus and Silvaco TCAD packages to produce converging results of the basic sensor properties (that also include uncertainties). This would lead to a more complete understanding and control over the simulation process, and potentially to merging defect models between the packages. It should not be taken for granted that basic silicon properties are well-known~\cite{ATL-INDET-PUB-2018-001}.

The essential objective of the combined bulk and surface damage simulations is to stretch the validated defect modeling up to $\sim 2\times10^{16}$ n$_\textrm{\small eq}/$cm$^{2}$ to 
account for the fluences of the pixel and 3D-columnar pixel detectors positioned closest to the vertex at the HL-LHC, as well as for sensors utilized in other extreme-fluence future 
colliders (e.g. $e^+$/$e^-$ and hadron colliders). Presently the validation limit of most of the peer-reviewed published models is about $1\times10^{15}$ n$_{\textrm{\small eq}}/$cm$^{2}$, 
with the exception of separate characteristics like LGAD leakage current in \cite{Croci:2022ygd} and CCE of 300-$\upmu$m-thick strip sensors and 3D-pixel sensors in \cite{Passeri2016} and~\cite{Dawson:2764325}, respectively, in the $10^{16}$ n$_{\textrm{\small eq}}/$cm$^{2}$ fluence region. 

All these TCAD-simulation R\&D branches will require significant amount of effort (time, budget, personnel, etc.) and will be difficult to realize without a dedicated and well-organized collaboration.
%

\section*{\label{sec::acknowledgments}Acknowledgments}

BN was supported by the Department of Energy, Office of Science under contract number DE-AC02-05CH11231. XS was supported by the National Natural Science Foundation of China (No. 1196114014). 

\appendix
\section{TCAD mathematical models}\label{TCADmathModels} 

TCAD software suites are designed to simulate charge transport and electrical polarization of semiconductor devices using a semi classical approach to electronic transport. The essential set of equations, \cite{Fu2012}, takes into account the Poisson equation, \ref{eq:poissoneq} for the electric field and charges densities, the current continuity, eqs. \ref{eq:continuitye}, \ref{eq:continuityh} and the trap occupation dynamics, eqs.\ref{eq:occupancy} to \ref{eq:htrapdynamics}:
\begin{eqnarray}
\nabla \cdot(\epsilon \nabla \phi)&=&-q(p-n+N_D(1-f^n_D) -N_A f^n_A) - q\sum_j N_{tj}(\delta_j-f^n_{tj})
\label{eq:poissoneq}\\
\frac{\partial n}{\partial t} - N_D \frac{\partial f^n_D}{\partial t}&=&\big(G_{net,n}-\sum_j R_n^{tj} - R_{au}\big)+\frac{1}{q}\nabla \cdot \overrightarrow{J_n}
\label{eq:continuitye}\\
\frac{\partial p}{\partial t} + N_A \frac{\partial f^n_A}{\partial t}&=&\big(G_{net,p}-\sum_j R_p^{tj}-R_{au}\big)-\frac{1}{q}\nabla \cdot \overrightarrow{J_p}
\label{eq:continuityh}\\
N_{tj}\frac{\partial f^n_{tj}}{\partial t}&=& R_n^{tj}-R_p^{tj}\qquad \textrm{for each trap j, where:}
\label{eq:occupancy}\\
R_n^{tj}&=&c_{nj}n N_{tj}(1-f^n_{tj})-e_{nj}N_{tj}f^n_{tj}
\label{eq:etrapdynamics}\\
R_p^{tj}&=&c_{pj}p N_{tj}f^n_{tj}-e_{pj}N_{tj}(1-f^n_{tj})
\label{eq:htrapdynamics}
\end{eqnarray}

\noindent The summation term in Eq.(\ref{eq:poissoneq}) is  $\rho_{trap}$, the trapped net charge, where $\delta_j=1$ if the trap j is of donor type, $\delta_j=0$ if the trap j is of acceptor type. In Eqs.(\ref{eq:etrapdynamics}, \ref{eq:htrapdynamics}) $c_{nj},\,c_{pj}$ are the electron/hole capture terms and $e_{nj},\,e_{pj}$ are the electron/hole emission terms for trap $j$. 
$f^n_{tj}$ means the electron occupation fraction of trap $j$ and $f^n_D$, $f^n_A$ are the occupancies for dopant donors and acceptors. $G_{net,n},\,G_{net,p}$ are the net generation rates for electrons and holes, including optical, radiation, impact ionization and other available generation mechanisms. Correspondingly, $R_{au}$ is the Auger recombination and $R_{n}^{tj},\,R_{p}^{tj}$ are the electron and hole recombination rates for the trap $j$.  

Different TCAD software providers offers slightly different set of equations for the currents but all of them implement the drift-diffusion (DD) currents for $\overrightarrow J_n$ and $\overrightarrow J_p$. DD is appropriate for simulation of semiconductor particle detectors at constant temperature. The semiconductor is not in equilibrium so it is necessary to define Quasi-Fermi potentials, $\Phi_{n},\Phi_{p}$ \cite{grundmann}. The DD currents definition is, \cite{Chang1988}:
\begin{eqnarray}
\overrightarrow J_n&=&-nq\mu_n\nabla\Phi_n=\mu_n(n\nabla E_c -\frac{3}{2}nKT\nabla ln\, m_n)+D_n(\nabla n - n\,\nabla ln\,\gamma_n)
\label{eq:jpchang}\\
\overrightarrow J_p&=&-pq\mu_p\nabla\Phi_p=\mu_p(n\nabla E_v -\frac{3}{2}nKT\nabla ln\, m_p)-D_p(\nabla p - p\,\nabla ln\,\gamma_p)
\label{eq:jnchang}
\end{eqnarray}
\noindent that considers the contributions from drift, diffusion and also spatial variations of the effective mass and Fermi-Dirac statistics. 

\bibliographystyle{JHEP}
\bibliography{main}

\end{document}